\documentclass[10pt, letterpaper]{article}
\usepackage{setspace}
\usepackage{fullpage} 

\usepackage{graphicx}
\usepackage{epsfig}
\usepackage{alltt}
\usepackage{breakcites}

\newcommand{\fullonly}[1]{#1}
\newcommand{\shortonly}[1]{}

\newcommand{\noise}[1]{#1}
\newcommand{\nonoise}[1]{}

\newcommand{\lncsonly}[1]{}
\newcommand{\articleonly}[1]{#1}

\newcounter{rulecounter}[section]

\newcommand{\comment}[1]{}

\usepackage{epsfig}
\usepackage{url}
\usepackage{ifpdf}
\usepackage{graphicx}

\newcommand{\tuple}[1]{\langle #1\rangle}
\newcommand{\mean}[1]{\left[ \! \left[ #1 \right]\! \right]}
\newcommand{\setc}[2]{\{#1 \;|\; #2\}}
\newcommand{\intersect}{\cap}

\newcommand{\Op}{{\it Op}}
\newcommand{\uid}{{\rm uid}}
\newcommand{\rid}{{\rm rid}}
\newcommand{\vals}{{\rm Val}_{\rm s}}
\newcommand{\valm}{{\rm Val}_{\rm m}}
\newcommand{\up}{{\it UP}}
\newcommand{\myparagraph}[1]{\paragraph*{#1.}}
\newcommand{\attrib}{{\rm attr}}

\newcommand{\union}{\cup}
\newcommand{\UNION}{\bigcup}
\newcommand{\powerset}{{\rm Set}}
\newcommand{\Rho}{{\it Rules}}

\newcommand{\set}[1]{\{#1\}}
\newcommand{\dept}{{\rm dept}}
\newcommand{\cs}{{\rm CS}}

\newcommand{\position}{{\rm position}}
\newcommand{\grad}{{\rm grad}}
\newcommand{\ugrad}{{\rm ugrad}}
\newcommand{\courses}{{\rm courses}}
\newcommand{\uae}{{\rm uae}}
\newcommand{\rae}{{\rm rae}}
\newcommand{\ops}{{\rm ops}}
\newcommand{\con}{{\rm con}}
\newcommand{\uattr}{{\rm uAttr}}
\newcommand{\rattr}{{\rm rAttr}}
\newcommand{\uncovup}{{\it uncovUP}}

\newcommand{\cc}{{\it cc}}
\newcommand{\gen}{{\it gen}}
\newcommand{\isempty}{{\rm isEmpty}}
\newcommand{\add}{{\rm add}}
\newcommand{\addall}{{\rm addAll}}
\newcommand{\removeall}{{\rm removeAll}}
\newcommand{\removeElt}{{\rm remove}}
\newcommand{\Qpol}{Q_{\rm pol}}

\newcounter{lnum}

\newcommand{\uad}{d_{\rm u}}
\newcommand{\rad}{d_{\rm r}}
\newcommand{\Au}{A_{\rm u}}
\newcommand{\Aus}{A_{\rm u,1}}
\newcommand{\Aum}{A_{\rm u,m}}
\newcommand{\Ar}{A_{\rm r}}
\newcommand{\Ars}{A_{\rm r,1}}
\newcommand{\Arm}{A_{\rm r,m}}
\newcommand{\Aunrm}{A_{\rm unrm}}

\newcommand{\aus}{a_{\rm u,1}}
\newcommand{\aum}{a_{\rm u,m}}

\newcommand{\ars}{a_{\rm r,1}}
\newcommand{\arm}{a_{\rm r,m}}
\newcommand{\eu}{e_{\rm u}}
\newcommand{\er}{e_{\rm r}}
\newcommand{\su}{s_{\rm u}}
\newcommand{\so}{s_{\rm o}}
\newcommand{\supseteqin}{\mathrel{\mathord{\supseteq}\mathord{\in}}}
\newcommand{\rhobest}{\rho_{\rm best}}
\newcommand{\rhomerge}{\rho_{\rm mrg}}
\newcommand{\wkset}{{\it workSet}}
\newcommand{\stringlit}[1]{\mbox{\tt "}{\rm #1}\mbox{\tt "}}

\newcommand{\ta}{{\it ta}}

\newcommand{\redrules}{{\it redun}}

\newcommand{\true}{{\rm true}}
\newcommand{\false}{{\rm false}}
\newcommand{\freq}{{\rm freq}}

\newcommand{\pigen}{\pi_{\rm gen}}

\newcommand{\Nrule}{N_{\rm rule}}

\newcommand{\com}[1]{{\it #1}}

\newcommand{\function}{{\bf function}}
\newcommand{\ifstmt}{{\bf if}}
\newcommand{\elsestmt}{{\bf else}}
\newcommand{\eifstmt}{{\bf end if}}
\newcommand{\forloop}{{\bf for}}

\newcommand{\whileloop}{{\bf while}}
\newcommand{\eforloop}{{\bf end for}}
\newcommand{\ewhileloop}{{\bf end while}}

\newcommand{\forloopin}{{\bf in}}
\newcommand{\forloopto}{{\bf to}}
\newcommand{\return}{{\bf return}}

\newcommand{\computeuae}{{\rm computeUAE}}
\newcommand{\computerae}{{\rm computeRAE}}
\newcommand{\addcandidaterule}{{\rm addCandRule}}
\newcommand{\candidateconstraint}{{\rm candidateConstraint}}
\newcommand{\elimattrib}{{\rm elimAttribute}}
\newcommand{\elimconstraints}{{\rm elimConstraints}}

\newcommand{\elimelements}{{\rm elimElements}}
\newcommand{\elimconjuncts}{{\rm elimConjuncts}}
\newcommand{\elimconjunctshelper}{{\rm elimConjunctsHelper}}
\newcommand{\elimoverlapVal}{{\rm elimOverlapVal}}
\newcommand{\elimoverlapOp}{{\rm elimOverlapOp}}
\newcommand{\elimredundantsets}{{\rm elimRedundantSets}}
\newcommand{\generalizerule}{{\rm generalizeRule}}
\newcommand{\maxConjunctSz}{{\rm maxConjunctSz}}
\newcommand{\mergerules}{{\rm mergeRules}}

\newcommand{\Qrul}{Q_{\rm rul}}
\newcommand{\fwQrul}{\Qrul^{\rm freq}}
\newcommand{\QrulILP}{\Qrul^{\rm ILP}}
\newcommand{\simplifyrules}{{\rm simplifyRules}}
\newcommand{\wsc}{{\rm WSC}}

\newcommand{\freqL}{\hyperlink{freq}{\freq}}
\newcommand{\addcandidateruleL}{\hyperlink{addcandidaterule}{\addcandidaterule}}
\newcommand{\computeuaeL}{\hyperlink{computeuae}{\computeuae}}
\newcommand{\computeraeL}{\hyperlink{computerae}{\computerae}}
\newcommand{\candidateconstraintL}{\hyperlink{candidateconstraint}{\candidateconstraint}}
\newcommand{\elimattribL}{\hyperlink{elimattrib}{\elimattrib}}
\newcommand{\elimconstraintsL}{\hyperlink{elimconstraints}{\elimconstraints}}

\newcommand{\elimelementsL}{\hyperlink{elimelements}{\elimelements}}
\newcommand{\elimconjunctsL}{\hyperlink{elimconjuncts}{\elimconjuncts}}
\newcommand{\elimconjunctshelperL}{\hyperlink{elimconjunctshelper}{\elimconjunctshelper}}
\newcommand{\elimoverlapValL}{\hyperlink{elimoverlapVal}{\elimoverlapVal}}
\newcommand{\elimoverlapOpL}{\hyperlink{elimoverlapOp}{\elimoverlapOp}}
\newcommand{\elimredundantsetsL}{\hyperlink{elimredundantsets}{\elimredundantsets}}
\newcommand{\generalizeruleL}{\hyperlink{generalizerule}{\generalizerule}}
\newcommand{\maxConjunctSzL}{\hyperlink{maxConjunctSz}{\maxConjunctSz}}
\newcommand{\mergerulesL}{\hyperlink{mergerules}{\mergerules}}

\newcommand{\QrulL}{\hyperlink{Qrul}{\Qrul}}
\newcommand{\fwQrulL}{\hyperlink{fwQrul}{\fwQrul}}

\newcommand{\simplifyrulesL}{\hyperlink{simplifyrules}{\simplifyrules}}

\newcommand{\Copy}{{\rm copy}}

\newcommand{\synSim}{S_{\rm syn}}
\newcommand{\uaeSim}{\synSim^{\rm u}}
\newcommand{\raeSim}{\synSim^{\rm r}}
\newcommand{\rules}{{\rm rules}}

\newcommand{\pcomp}{{\it pcomp}}

\newcommand{\userperm}{{\rm userPerm}}

\newcommand{\Prule}{P_{\rm rule}}
\newcommand{\PuserRes}{P_{\rm ur}}
\newcommand{\Pop}{P_{\rm op}}
\newcommand{\Puser}{P_{\rm user}}
\newcommand{\Pres}{P_{\rm res}}
\newcommand{\PopSimp}{P_{\rm op}'}

\begin{document}

\title{Mining Attribute-Based Access Control Policies from Logs\thanks{This
    material is based upon work supported in part by %
    NSF under Grant CNS-0831298. 
}
}

\articleonly{\author{Zhongyuan Xu and Scott~D.~Stoller\\
    Department of Computer Science, Stony Brook University, USA}}

\lncsonly{\author{Zhongyuan Xu \and Scott~D.~Stoller}
  \institute{Department of Computer Science, Stony Brook University, USA}}


\maketitle

\begin{abstract}
  Attribute-based access control (ABAC) provides a high level of
  flexibility that promotes security and information sharing.
  ABAC policy mining algorithms have potential to significantly reduce the
  cost of migration to ABAC, by partially automating the development of an
  ABAC policy from information about the existing access-control policy and
  attribute data.  This paper presents an algorithm for mining ABAC
  policies from operation logs and attribute data.  To the best of our
  knowledge, it is the first algorithm for this problem.
\end{abstract}




\section{Introduction}
\label{sec:intro}

ABAC is becoming increasingly important as security policies become more
dynamic and more complex.  In industry, more and more products support
ABAC, using a standardized ABAC language such as XACML\fullonly{
  \cite{XACML}} or a vendor-specific ABAC language.  In government, the
Federal Chief Information Officer Council called out ABAC as a recommended
access control model
\shortonly{\cite{FEDCIO11short,NIST13ABAC-URLshort}}\fullonly{\cite{FEDCIO11,NIST13ABAC-URL}}.
ABAC allows ``an unprecedented amount of flexibility and security while
promoting information sharing between diverse and often disparate
organizations''
\shortonly{\cite{NIST13ABAC-URLshort}}\fullonly{\cite{NIST13ABAC-URL}}.
ABAC overcomes some of the problems associated with RBAC, notably role
explosion
\shortonly{\cite{NIST13ABAC-URLshort}}\fullonly{\cite{NIST13ABAC-URL},
  which makes RBAC policies large and hard to manage}.

ABAC promises long-term cost savings through reduced management effort, but
manual development of an initial policy can be difficult\fullonly{
  \cite{beckerle13formal}} and expensive \shortonly{\cite{NIST13ABAC-URLshort}}\fullonly{\cite{NIST13ABAC-URL}}.  {\em
  Policy mining} algorithms promise to drastically reduce the cost of
migrating to ABAC, by partially automating the process.

Role mining, i.e., mining of RBAC policies, is an active research
area\fullonly{ (e.g.,
  \cite{conf/sacmat/kuhlm03,conf/sacmat/schle05,vaidya06roleminer,vaidya07finding,zhang07role,colantonio08cost-driven,frank08class,guo08role,lu08optimal,vaidya08migrating,colantonio09formal,molloy09evaluating,conf/sacmat/FrankBB10,ma10role,molloy10mining,molloy10noisy,takabi10stateminer,vaidya10role,colantonio12decomposition,lu12constraint,molloy12generative,verde12role,xu12algorithms,frank13role,mitra13toward,xu13mining,zhang13evolving})}
and a currently relatively small (about \$70 million) but rapidly growing
commercial market segment \cite{hachana12role}.\fullonly{ Some form of role
  mining is supported by security products from companies of all sizes,
  from IT giants (e.g., IBM Tivoli Access Manager, Oracle Role Manager, CA
  Technologies GovernanceMinder) to small start-ups (e.g., Bay31 Role
  Designer).}  In contrast, there is, so far, relatively little work on
ABAC policy mining.  We recently developed an algorithm to mine an ABAC
policy from an ACL policy or RBAC policy \cite{xu13miningABAC}\fullonly{;
  to the best of our knowledge, it is the first algorithm for that problem}.

However, an ACL policy or RBAC policy might not be available, e.g., if the
current access control policy is encoded in a program or is not enforced by
a computerized access control mechanism.  An alternative source of
information about the current access control policy is operation logs, or
``logs'' for short.  Many software systems produce logs, e.g., for
auditing, accounting, and accountability purposes.  Molloy, Park, and Chari
proposed the idea of mining policies from logs and developed algorithms for
mining RBAC policies from logs \shortonly{\cite{molloy12generativeShort}}\fullonly{\cite{molloy12generative}}.

The main challenge is that logs generally provide incomplete information
about entitlements (i.e., granted permissions).  Specifically, logs provide
only a lower bound on the entitlements.  Therefore, the generated policy
should be allowed to include {\em over-assignments}, i.e., entitlements not
reflected in the logs.


This paper presents an algorithm for mining ABAC policies from logs and
attribute data.  To the best of our knowledge, it is the first algorithm
for this problem.  It is based on our algorithm for mining ABAC policies
from ACLs \cite{xu13miningABAC}.  At a high level, the algorithm works as
follows.  It iterates over tuples in the user-permission relation extracted
from the log, uses selected tuples as seeds for constructing candidate
rules, and attempts to generalize each candidate rule to cover additional
tuples in the user-permission relation by replacing conjuncts in attribute
expressions with constraints.  After constructing candidate rules that
together cover the entire user-permission relation, it attempts to improve
the policy by merging and simplifying candidate rules.  Finally, it selects
the highest-quality candidate rules for inclusion in the generated policy.

Several changes are needed to our algorithm for mining ABAC policies from
ACLs to adapt it to mining from logs.  When the algorithm generalizes,
merges, or simplifies rules, it discards candidate rules that are invalid,
i.e., that produce over-assignments.  We modify those parts of the
algorithm to consider those candidate rules, because, as discussed above,
over-assignments must be permitted.  To evaluate those candidate rules, we
introduce generalized notions of rule quality and policy quality that
quantify a trade-off between the number of over-assignments and other
aspects of quality.  We consider a metric that includes the normalized
number of over-assignments in a weighted sum, a frequency-sensitive variant
that assigns higher quality to rules that cover more frequently used
entitlements, along the lines of
\shortonly{\cite{molloy12generativeShort}}\fullonly{\cite{molloy12generative}},
and a metric based on a theory quality metric in inductive logic
programming \shortonly{\cite{muggleton95inverse,muggleton01cprogol}}\fullonly{\cite{muggleton95inverse,muggleton96learning,muggleton01cprogol}}.\noise{ We also
  modify Xu and Stoller's approach to noise detection to take usage
  frequency into account.}

ABAC policy mining is similar to inductive logic programming (ILP), which
learns logic-programming rules from facts.  Mining ABAC policies from logs
and attribute data is similar to ILP algorithms for learning from positive
examples, because those algorithms allow the learned rules to imply more
than the given facts (i.e., in our terminology, to have over-assignments).
We implemented a translation from ABAC policy mining to Progol
\shortonly{\cite{muggleton01cprogol}}\fullonly{\cite{muggleton96learning,muggleton01cprogol}}, a well-known ILP system.

We evaluated our algorithm and the ILP-based approach on some relatively
small but non-trivial handwritten case studies and on synthetic ABAC
policies.  The results demonstrate our algorithm's effectiveness even when
the log reflects only a fraction of the entitlements.\shortonly{ Although
  the original (desired) ABAC policy is not reconstructed perfectly from
  the log, the mined policy is sufficiently similar to it that the mined
  policy would be very useful as a starting point for policy administrators
  tasked with developing that ABAC policy.}



\section{ABAC policy language}
\label{sec:language}

This section presents the ABAC policy language used in our work.  It is
adopted from \cite{xu13miningABAC}.
\fullonly{For simplicity and concreteness, we}\shortonly{We} consider a
specific ABAC policy language, but our approach is general and can be
adapted to other ABAC policy languages. Our ABAC policy language contains
the common ABAC policy language constructs, except arithmetic inequalities
and negation, which are left for future work.

\fullonly{ABAC policies refer to attributes of users and resources.}  Given
a set $U$ of users and a set $\Au$ of user attributes, user attribute data
is represented by a function $\uad$ such that $\uad(u,a)$ is the value of
attribute $a$ for user $u$.  There is a distinguished user attribute $\uid$
that has a unique value for each user.  Similarly, given a set $R$ of
resources and a set $\Ar$ of resource attributes, resource attribute data
is represented by a function $\rad$ such that $\rad(r,a)$ is the value of
attribute $a$ for resource $r$.  There is a distinguished resource
attribute $\rid$ that has a unique value for each resource.  We assume the
set $\Au$ of user attributes can be partitioned into a set $\Aus$ of {\it
  single-valued user attributes} which have atomic values, and a set $\Aum$
of {\it multi-valued user attributes} whose values are sets of atomic
values.  Similarly, we assume the set $\Ar$ of resource attributes can be
partitioned into a set $\Ars$ of {\it single-valued resource attributes}
and a set of $\Arm$ of {\it multi-valued resource attributes}.\fullonly{
  Let $\vals$ be the set of possible atomic values of attributes.  We
  assume $\vals$ includes a distinguished value $\bot$ used to indicate
  that an attribute's value is unknown.  The set of possible values of
  multi-valued attributes is
  $\valm=\powerset(\vals\setminus\set{\bot})\union\bot$, where
  $\powerset(S)$ is the powerset of set $S$.}\shortonly{ We assume there is
  a distinguished atomic value $\bot$ used to indicate that an attribute's
  value is unknown.}

\fullonly{Attribute expressions are used to express the sets of users and
  resources to which a rule applies.}  A {\em user-attribute expression}
(UAE) is a function $e$ such that, for each user attribute $a$, $e(a)$ is
either the special value $\top$, indicating that $e$ imposes no constraint
on the value of $a$, or a set (interpreted as a disjunction) of possible
values of $a$ excluding $\bot$\fullonly{ (in other words, a subset of
  $\vals\setminus\set{\bot}$ or $\valm\setminus\set{\bot}$, depending on
  whether $a$ is single-valued or multi-valued)}.  We refer to $e(a)$ as
the {\em conjunct} for\fullonly{ attribute} $a$.  A\fullonly{
  user-attribute expression}\shortonly{ UAE} $e$ {\em uses} attribute $a$
if $e(a)\ne \top$.  Let $\attrib(e)$ denote the set of attributes used by
$e$.


A user $u$ {\em satisfies} a\fullonly{ user-attribute
  expression}\shortonly{ UAE} $e$, denoted $u \models e$, iff $(\forall
a\in \Aus.\, e(a)=\top \lor \exists v\in e(a).\, \uad(u,a)=v)$ and
$(\forall a\in \Aum.\, e(a)=\top \lor \exists v\in e(a).\,
\uad(u,a)\supseteq v)$.  For multi-valued attributes, we use the condition
$\uad(u,a)\supseteq v$ instead of $\uad(u,a)=v$ because elements of a
multi-valued user attribute typically represent some type of capabilities
of a user, so using $\supseteq$ expresses that the user has the specified
capabilities and possibly more.\fullonly{

}\shortonly{ }For example, suppose $\Aus =\{\dept, \position\}$ and $\Aum
=\{\courses\}$.  The function $e_1$ with $e_1(\dept) = \{\cs\}$ and
$e_1(\position) = \{\grad,\ugrad\}$ and $e_1(\courses)=\set{\set{{\rm
      CS101}, {\rm CS102}}}$ is a user-attribute expression satisfied by
users in the $\cs$ department who are either graduate or undergraduate
students and whose courses include CS101 and
CS102. 

\shortonly{In examples, we may write attribute expressions with a
  logic-based syntax, for readability.}\fullonly{We introduce a concrete
  syntax for attribute expressions, for improved readability in examples.
  We write a user attribute expression as a conjunction of the conjuncts
  not equal to $\top$.  Suppose $e(a)\ne\top$.  Let $v=e(a)$.  When $a$ is
  single-valued, we write the conjunct for $a$ as $a \in v$; as syntactic
  sugar, if $v$ is a singleton set $\set{s}$, we may write the conjunct as
  $a=s$.  When $a$ is multi-valued, we write the conjunct for $a$ as $a
  \supseteqin v$ (indicating that $a$ is a superset of an element of $v$);
  as syntactic sugar, if $v$ is a singleton set $\set{s}$, we may write the
  conjunct as $a\supseteq s$.}  For example, the above expression $e_1$ may
be written as $\dept=\cs \land \position \in \set{\ugrad,\grad} \land
\courses \supseteq \set{{\rm CS101}, {\rm CS102}}$.  For an example that
uses $\supseteqin$, the expression $e_2$ that is the same as $e_1$ except
with $e_2(\courses)=\set{\set{{\rm CS101}}, \set{{\rm CS102}}}$ may be
written as $\dept=\cs \land \position \in \set{\ugrad,\grad} \land \courses
\supseteqin \set{\set{{\rm CS101}}, \set{{\rm CS102}}}$, and is satisfied
by graduate or undergraduate students in the $\cs$ department whose courses
include either CS101 or CS102.

The {\em meaning} of a user-attribute expression $e$, denoted $\mean{e}_U$,
is the set of users in $U$ that satisfy it\fullonly{: $\mean{e}_U = \setc{u
    \in U}{u \models e}$}.  User attribute data is an implicit argument to
$\mean{e}_U$.  We say that $e$ {\em characterizes} the set $\mean{e}_U$.

%
%

A {\em resource-attribute expression} (RAE) is defined similarly, except
using the set $\Ar$ of resource attributes instead of the set $\Au$ of user
attributes.  The semantics of RAEs is defined similarly to the semantics of
UAEs, except simply using equality, not $\supseteq$, in the condition for
multi-valued attributes in the definition of ``satisfies'', because we do
not interpret elements of multi-valued resource attributes specially (e.g.,
as capabilities).

Constraints express relationships between users and resources.  An {\em
  atomic constraint} is a formula $f$ of the form $\aum\supseteq\arm$,
$\aum\ni\ars$, or $\aus=\ars$, where $\aus\in\Aus$, $\aum\in\Aum$,
$\ars\in\Ars$, and $\arm\in\Arm$.  The first two forms express that user
attributes contain specified values.  This is a common type of constraint,
because user attributes typically represent some type of capabilities of a
user.\fullonly{ Other forms of atomic constraint are possible (e.g.,
  $\aum\subseteq\arm$) but less common, so we leave them for future work.}
Let $\uattr(f)$ and $\rattr(f)$ refer to the user attribute and resource
attribute, respectively, used in $f$.  User $u$ and resource $r$ {\em
  satisfy} an atomic constraint $f$, denoted $\tuple{u, r} \models f$, if
$\uad(u,\uattr(f))\ne\bot$ and $\rad(u,\rattr(f))\ne\bot$ and formula $f$
holds when the values $\uad(u,\uattr(f))$ and $\rad(u,\rattr(f))$ are
substituted in it.

A {\em constraint} is a set (interpreted as a conjunction) of atomic
constraints.  User $u$ and resource $r$ {\em satisfy} a constraint $c$,
denoted $\tuple{u, r} \models c$, if they satisfy every atomic constraint
in $c$.  In examples, we write constraints as conjunctions instead of sets.
For example, the constraint ``{\rm specialties} $\supseteq$ {\rm topics}
$\,\land\,$ {\rm teams} $\ni$ {\rm treatingTeam}'' is satisfied by user $u$ and
resource $r$ if the user's specialties include all of the topics associated
with the resource, and the set of teams associated with the user contains
the treatingTeam associated with the resource.

A {\em user-permission tuple} is a tuple $\tuple{u, r, o}$ containing a
user, a resource, and an operation.  This tuple means that user $u$ has
permission to perform operation $o$ on resource $r$.  A {\em
  user-permission relation} is a set of such tuples.

A {\em rule} is a tuple $\tuple{\eu, \er, O, c}$, where $\eu$ is a
user-attribute expression, $\er$ is a resource-attribute expression, $O$ is
a set of operations, and $c$ is a constraint.  For a rule $\rho=\tuple{\eu,
  \er, O, c}$, let $\uae(\rho)=\eu$, $\rae(\rho)=\er$, $\ops(\rho)=O$, and
$\con(\rho)=c$.  For example, the rule $\langle$true, type=task $\land$
proprietary=false, \{read, request\}, projects $\ni$ project $\land$
expertise $\supseteq$ expertise$\rangle$ used in our project management
case study can be interpreted as ``A user working on a project can read and
request to work on a non-proprietary task whose required areas of expertise
are among his/her areas of expertise.''  User $u$, resource $r$, and
operation $o$ {\em satisfy} a rule $\rho$, denoted $\tuple{u, r, o} \models
\rho$, if $u\models \uae(\rho) \land r\models\rae(\rho) \land o \in
\ops(\rho) \land \tuple{u,r}\models \con(\rho)$.


An {\em ABAC policy} is a tuple $\tuple{U, R, \Op, \Au, \Ar, \uad, \rad,
  \Rho}$, where $U$, $R$, $\Au$, $\Ar$, $\uad$, and $\rad$ are as described
above, $\Op$ is a set of operations, and $\Rho$ is a set of rules.

The user-permission relation induced by a rule $\rho$ is $\mean{\rho} =
\setc{\tuple{u,r,o} \in U\times R\times \Op}{\tuple{u,r,o}\models \rho}$.
Note that $U$, $R$, $\uad$, and $\rad$ are implicit arguments to
$\mean{\rho}$.

The user-permission relation induced by a policy $\pi$ with the above form
is $\mean{\pi} = \UNION_{\rho \in \Rho} \mean{\rho}$.



\section{Problem Definition}
\label{sec:problem}

An {\em operation log entry} $e$ is a tuple $\tuple{u,r,o,t}$ where $u\in
U$ is a user, $r\in R$ is a resource, $o\in Op$ is an operation, and $t$ is
a timestamp.  An {\em operation log} is a sequence of operation log
entries.  The user-permission relation induced by an operation log $L$ is
$\up(L) = \setc{\tuple{u,r,o}}{\exists t. \; \tuple{u,r,o,t}\in L}$.

The input to the {\em ABAC-from-logs policy mining problem} is a tuple $I =
\tuple{U, R, \Op, \Au, \Ar, \uad, \rad, L}$, where $U$ is a
set of users, $R$ is a set of resources, $\Op$ is a set of operations,
$\Au$ is a set of user attributes, $\Ar$ is a set of resource attributes,
$\uad$ is user attribute data, $\rad$ is resource attribute data, and $L$
is an operation log, such that the users, resources, and operations that
appear in $L$ are subsets of $U$, $R$, and $\Op$, respectively.  The goal
of the problem is to find a set of rules $\Rho$ such that the ABAC policy
$\pi=\tuple{U, R, \Op, \Au, \Ar, \uad, \rad, \Rho}$ maximizes a suitable
policy quality metric.


The policy quality metric should reflect the size and meaning of the
policy.  Size is measured by {\em weighted structural complexity} (WSC)
\cite{molloy10mining}, and smaller policies are considered to have higher
quality.  This is consistent with usability studies of access control
rules, which conclude that more concise policies are more
manageable\fullonly{ \cite{beckerle13formal}}.  Informally, the WSC of an
ABAC policy is a weighted sum of the number of elements in the policy.
Specifically, the WSC of an attribute expression is the number of atomic
values that appear in it, the WSC of an operation set is the number of
operations in it, the WSC of a constraint is the number of atomic
constraints in it, and the WSC of a rule is a weighted sum of the WSCs of
its components, namely, $\hypertarget{wscRule}{\wsc(\tuple{\eu, \er, O,
    c})} = w_1\wsc(\eu) + w_2\wsc(\er) + w_3\wsc(O) + w_4\wsc(c)$, where
the $w_i$ are user-specified weights.  The WSC of a set of rules is the sum
of the WSCs of its members.

The meaning $\mean{\pi}$ of the ABAC policy is taken into account by
considering the differences from $\up(L)$, which consist of
over-assignments and under-assignments.  The over-assignments are
$\mean{\pi}\setminus\up(L)$.  The under-assignments are
$\up(L)\setminus\mean{\pi}$.  Since logs provide only a lower-bound on the
actual user-permission relation (a.k.a entitlements), it is necessary to
allow some over-assignments, but not too many.  Allowing under-assignments
is beneficial if the logs might contain noise, in the form of log entries
representing uses of permissions that should not be granted, because it
reduces the amount of such noise that gets propagated into the mined
policy\fullonly{, and it improves the stability of the generated
  policy}\noise{.}\nonoise{; consideration of noise is left for future work}.
We define a policy quality metric that is a weighted sum of these aspects:
\begin{equation}
  \label{eq:Qpolicy}
  \Qpol(\pi, L) = \wsc(\pi) + w_o\,|\mean{\pi}\setminus\up(L)|\,/\,|U|\noise{ + w_u|\up(L)\setminus\mean{\pi}|_L}
\end{equation}
\shortonly{where the {\em policy over-assignment weight} $w_o$ is a
  user-specified weight for over-assignments, }%
\noise{where the {\em policy over-assignment weight} $w_o$ and the {\em
    under-assignment weight} $w_u$ are user-specified weights for
  over-assignments and under-assignments, respectively, }%
and for a set $S$ of user-permission tuples, the frequency-weighted size of
$S$ with respect to log $L$ is $|S|_L = \sum_{\tuple{u,r,o}\in S}
\freq(\tuple{u,r,o}, L)$, where the relative frequency of a user-permission
tuple in a log is given by the {\em frequency function}
$\hypertarget{freq}{\freq(\tuple{u,r,o}, L)}=|\setc{e \in L}{\userperm(e)=\tuple{u,r,o}}|\,/\,|L|$,
where the user-permission part of a log entry is given by
$\userperm(\tuple{u,r,o,t})=\tuple{u,r,o}$.\fullonly{ Using the relative
  frequency, instead of the number of occurrences, of a user-permission
  tuple as the weight in the definition of frequency-weighted size allows
  $w_u$ to be chosen independent of the duration of monitoring and the
  number of monitored users; for example, if the duration of monitoring or
  the number of monitored users is doubled, and the distribution of their
  behaviors and the value of $w_u$ are kept unchanged, then the policy
  quality is unchanged.  Similarly, dividing the number of over-assignments
  by $|U|$ allows $w_o$ to be chosen independent of the number of users
  (and the duration of monitoring).}

For simplicity, our presentation of the problem and algorithm assume that
attribute data does not change during the time covered by the log.
Accommodating changes to attribute data is not difficult.  It mainly
requires re-defining the notions of policy quality and rule quality
(introduced in Section \ref{sec:algorithm}) to be based on the set of log
entries covered by a rule, denoted $\mean{\rho}_{\rm LE}$, rather than
$\mean{\rho}$.  The definition of $\mean{\rho}_{\rm LE}$ is similar to the
definition of $\mean{\rho}$, except that, when determining whether a log
entry is in $\mean{\rho}_{\rm LE}$, the attribute data in effect at the
time of the log entry is used.


\section{Algorithm}
\label{sec:algorithm}

Our algorithm is based on the algorithm for mining ABAC policies from ACLs
and attribute data in \cite{xu13miningABAC}.  Our algorithm does not take
the order of log entries into account, so the log can be summarized by the
user-permission relation $\up_0$ induced by the log and the frequency
function $\freqL$, described in the penultimate paragraph of Section
\ref{sec:problem}.

Top-level pseudocode appears in Figure \ref{fig:alg}.  We refer to tuples
selected in the first statement of the first while loop as {\em seeds}.
\shortonly{The top-level pseudocode is explained by embedded comments.  It
  calls several functions, described next.}\fullonly{The top-level
  pseudocode is explained by embedded comments.  It calls several
  functions, described next.}  Function names hyperlink to pseudocode for
the function, if it is included in the paper, otherwise to the description
of the function.

The function $\addcandidateruleL(\su, s_r, \so, \cc, \uncovup, \Rho)$ in
Figure \ref{fig:addcandidaterule} first calls $\computeuaeL$ to compute a
user-attribute expression $\eu$ that characterizes $\su$, and
$\computeraeL$ to compute a resource-attribute expression $\er$ that
characterizes $s_r$. It then calls $\generalizeruleL(\rho, \cc, \uncovup,
\Rho)$ to generalize rule $\rho=\tuple{\eu, \er, \so, \emptyset}$ to
$\rho'$ and adds $\rho'$ to candidate rule set $\Rho$. The details of the
functions called by $\addcandidaterule$ are described next.

The function
\shortonly{\hypertarget{computeuae}{$\computeuaeL(s,U)$}}\fullonly{$\computeuaeL(s,U)$
  in Figure \ref{fig:computeuae}} computes a user-attribute expression
$\eu$ that characterizes the set $s$ of users.  Preference is given to
attribute expressions that do not use $\uid$, since attribute-based
policies are generally preferable to identity-based policies, even when
they have higher WSC, because attribute-based generalize better.\shortonly{
  Similarly, \hypertarget{computerae}{$\computerae(s,R)$} computes a
  resource-attribute expression that characterizes the set $s$ of
  resources.  Pseudocode for $\computeuaeL$ and $\computeraeL$ are
  omitted.}\fullonly{ After constructing a candidate expression $e$, it
  calls \hypertarget{elimredundantsets}{$\elimredundantsets(e)$}, which
  attempts to lower the WSC of $e$ by examining the conjunct for each
  multi-valued user attribute, and removing each set that is a superset of
  another set in the same conjunct; this leaves the meaning of the rule
  unchanged, because $\supseteq$ is used in the condition for multi-valued
  attributes in the semantics of user attribute expressions.
Pseudocode for $\elimredundantsetsL$ is straightforward and omitted.  The
expression $\eu$ returned by $\computeuaeL$ might not be minimum-sized
among expressions that characterize $s$: it is possible that some
attributes mapped to a set of values by $\eu$ can instead be mapped to
$\top$.  We defer minimization of $\eu$ until after the call to
$\generalizeruleL$ (described below), because minimizing $\eu$ before that
would reduce opportunities to find relations between values of user
attributes and resource attributes in $\generalizeruleL$.

The function \hypertarget{computerae}{$\computerae(s,R)$} computes a
resource-attribute expression that characterizes the set $s$ of resources.
The definition is the same as for $\computeuaeL$, except using resource
attributes instead of user attributes, and the call to
$\elimredundantsetsL$ is omitted.  Pseudocode for $\computeraeL$ is
omitted.

}
The function \hypertarget{candidateconstraint}{$\candidateconstraint(r,u)$}
returns a set containing all of the atomic constraints that hold between
resource $r$ and user $u$. Pseudocode for $\candidateconstraint$ is
straightforward and omitted.

The function $\generalizeruleL(\rho, \cc, \uncovup, \Rho)$ in Figure
\ref{fig:generalizerule} attempts to generalize rule $\rho$ by adding some
of the atomic constraints in $\cc$ to $\rho$ and eliminating the conjuncts
of the user attribute expression and/or the resource attribute expression
corresponding to the attributes used in those constraints, i.e., mapping
those attributes to $\top$.  We call a rule obtained in this way a {\em
  generalization} of $\rho$.  Such a rule is more general than $\rho$ in
the sense that it refers to relationships instead of specific values.
Also, the user-permission relation induced by a generalization of $\rho$ is
a superset of the user-permission relation induced by $\rho$.
$\generalizeruleL(\rho, \cc, \uncovup, \Rho)$ returns the
generalization $\rho'$ of $\rho$ with the best quality according to a given
rule quality metric.  Note that $\rho'$ may cover tuples that are already
covered (i.e., are in $\up$); in other words, our algorithm can generate
policies containing rules whose meanings overlap.

A {\em rule quality metric} is a function $\Qrul(\rho, \up)$ that maps a
rule $\rho$ to a totally-ordered set, with the ordering chosen so that
larger values indicate high quality.  The second argument $\up$ is a set of
user-permission tuples.
Our rule quality metric assigns higher quality to rules that cover more
currently uncovered user-permission tuples and have smaller size, with an
additional term that imposes a penalty for over-assignments, measured as a
fraction of the number of user-permission tuples covered by the rule, and
with a weight specified by a parameter $w'_o$, called the {\em rule
  over-assignment weight}.
\begin{displaymath}
  \hypertarget{Qrul}{\Qrul(\rho, \up)} = 
    \frac{|\mean{\rho} \intersect \up|}{|\rho|} \times
    (1 - \frac{w'_o\times|\mean{\rho}\setminus\up(L)|}{|\mean{\rho}|}).
\end{displaymath}
In $\generalizeruleL$, $\uncovup$ is the second argument to $\QrulL$, so
$\mean{\rho}\intersect\up$ is the set of user-permission tuples in $\up_0$
that are covered by $\rho$ and not covered by rules already in the policy.
The loop over $i$ near the end of the pseudocode for $\generalizeruleL$
considers all possibilities for the first atomic constraint in $\cc$ that
gets added to the constraint of $\rho$.  The function calls itself
recursively to determine the subsequent atomic constraints in $c$ that get
added to the constraint.

We also developed a frequency-sensitive variant of this rule quality
metric.  Let $\hypertarget{fwQrul}{\fwQrul}$ denote the frequency-weighted
variant of $\QrulL$, obtained by weighting each user-permission tuple by
its relative frequency (i.e., fraction of occurrences) in the log, similar
to the definition of $\lambda$-distance in \shortonly{\cite{molloy12generativeShort}}\fullonly{\cite{molloy12generative}}.
Specifically, the definition of $\fwQrulL$ is obtained from the definition
of $\QrulL$ by replacing $|\mean{\rho} \intersect \up|$ with $|\mean{\rho}
\intersect \up|_L$ (recall that $|\cdot|_L$ is defined in Section
\ref{sec:problem}).

We also developed a rule quality metric $\hypertarget{QrulILP}{\QrulILP}$
based closely on the theory quality metric for inductive logic programming
described in \cite{muggleton95inverse}.\shortonly{ Details of the
  definition are omitted to save space.}\fullonly{ Details of the
  definition appear in Appendix \ref{sec:rule-quality-ILP}.}


The function $\mergerulesL(\Rho)$ in Figure \ref{fig:mergerules} attempts
to improve the quality of $\Rho$ by removing redundant rules and merging
pairs of rules.  A rule $\rho$ in $\Rho$ is {\em redundant} if $\Rho$
contains another rule $\rho'$ such that every user-permission tuple in
$\up_0$ that is in $\mean{\rho}$ is also in $\mean{\rho'}$.  Informally,
rules $\rho_1$ and $\rho_2$ are merged by taking, for each attribute, the
union of the conjuncts in $\rho_1$ and $\rho_2$ for that attribute.  If
adding the resulting rule $\rhomerge$ to the policy and removing rules
(including $\rho_1$ and $\rho_2$) that become redundant improves policy
quality and $\rhomerge$ does not have over-assignments, then $\rhomerge$ is
added to $\Rho$, and the redundant rules are removed from $\Rho$.  As
optimizations (in the implementation, not reflected in the pseudocode),
meanings of rules are cached, and policy quality is computed incrementally.
$\mergerulesL(\Rho)$ updates its argument $\Rho$ in place, and it returns a
Boolean indicating whether any rules were merged.

\begin{figure}[tbp]
\begin{tabular}[t]{@{}l@{}}
// \com{$\Rho$ is the set of candidate rules}\\
$\Rho=\emptyset$\\
// \com{$\uncovup$ contains user-permission tuples}\\
// \com{in $\up_0$ that are not covered by $\Rho$}\\
$\uncovup=\up_0.\Copy()$\\
\whileloop\ $\neg\uncovup.\isempty()$\\
~~ // \com{Select an uncovered tuple as a ``seed''.}\\
~~   $\tuple{u,r,o}$ = some tuple in $\uncovup$\\
~~   $\cc = \candidateconstraintL(r, u)$\\
~~   // \com{$\su$ contains users with permission $\tuple{r,o}$}\\
~~   // \com{and that have the same candidate}\\
~~   // \com{constraint for $r$ as $u$}\\
~~   $\su = 
\begin{array}[t]{@{}l@{}}
  \{u'\in U \;|\; \tuple{u',r,o}\in \up_0\\
  \hspace*{2em} {} \land \candidateconstraintL(r,u')=\cc\}
\end{array}$\\
~~ $\addcandidaterule(\su, \set{r}, \set{o}, \cc, \uncovup, \Rho)$\\
~~ // \com{$\so$ is set of operations that $u$ can apply to $r$}\\
~~   $\so = \setc{o'\in \Op}{\tuple{u,r,o'}\in \up_0}$\\
~~ $\addcandidaterule(\set{u}, \set{r}, \so, \cc, \uncovup, \Rho)$\\
\ewhileloop
\end{tabular}
~
\begin{tabular}[t]{@{}l@{}}
// \com{Repeatedly merge and simplify}\\
// \com{rules, until this has no effect}\\
$\mergerulesL(\Rho)$\\
\whileloop\ $
\begin{array}[t]{@{}l@{}}
  \simplifyrulesL(\Rho) \\
  \mathrel{\&\&} \mergerulesL(\Rho)
\end{array}$\\
~~skip\\
\ewhileloop\\
// \com{Select high quality rules into $\Rho'$.}\\
$\Rho'$ = $\emptyset$\\
Repeatedly move highest-quality rule\\
from $\Rho$ to $\Rho'$ until\\
$\sum_{\rho\in\Rho'}\mean{\rho} \supseteq \up_0$, using \\
$\up_0\setminus\mean{\Rho'}$ as second argument to\\
$\Qrul$, and discarding a rule if it does\\
not cover any tuples in $\up_0$ currently\\
uncovered by $\Rho'$.\\
\return\ $\Rho'$
\end{tabular}
\caption{Policy mining algorithm.  The pseudocode starts in column 1 and
  continues in column 2.}
\label{fig:alg}
\end{figure}

\begin{figure}[tbp]
\begin{tabular}{@{}l@{}}
\hypertarget{addcandidaterule}{\function\ $\addcandidaterule(\su, s_r, \so, \cc, \uncovup, \Rho)$} \\
// \com{Construct a rule $\rho$ that covers \lncsonly{user-perm.\ }\articleonly{user-permission }tuples $\setc{\tuple{u,r,o}}{u\in \su\land r\in s_r \land o\in \so}$.}\\
$\eu = \computeuaeL(\su,U)$;\shortonly{~}\fullonly{\\}
$\er = \computeraeL(s_r,R)$;\shortonly{~}\fullonly{\\}
$\rho = \tuple{\eu, \er, \so, \emptyset}$\\
$\rho' = \generalizeruleL(\rho, \cc, \uncovup, \Rho)$;\shortonly{~}\fullonly{\\}
$\Rho.\add(\rho')$;\shortonly{~}\fullonly{\\}
$\uncovup.\removeall(\mean{\rho'})$
\end{tabular}
\caption{Compute a candidate rule $\rho'$ and add $\rho'$ to candidate rule set $\Rho$}
\label{fig:addcandidaterule}
\end{figure}

\fullonly{
\begin{figure}[tbp]
\begin{tabular}{@{}l@{}}
\hypertarget{computeuae}{\function\ $\computeuae(s,U)$}\\
// \com{First try to characterize set $s$ of users without using}\\
// \com{$\uid$. Use all other attributes which have known}\\
// \com{values for all users in $s$.}\\
$e = (
\begin{array}[t]{@{}l@{}}
  \lambda\, a \in \Au.\, \\
  a\!=\!\uid \,\lor\, (\exists\, u\in s.\, \uad(u,a)=\bot)
  \mathop{?} \top
  : \UNION_{u\in s} \uad(u,a))
\end{array}$\\
\ifstmt\ $\mean{e}_U \ne s$\\
~~~~ // \com{$\uid$ is needed to characterize $s$}\\
~~~~ $e(\uid)=\UNION_{u\in s} \uad(u,\uid)$\\
\eifstmt\\
$\elimredundantsets(e)$\\
\return\ $e$
\end{tabular}
\caption{Compute a user-attribute expression that characterizes set $s$ of
  users, where $U$ is the set of all users. }
\label{fig:computeuae}
\end{figure}
}

\begin{figure}[tbp]
\begin{tabular}[t]{@{}l@{}}
\hypertarget{generalizerule}{\function\ $\generalizerule(
  \begin{array}[t]{@{}l@{}}
    \rho, \cc, \uncovup, \\
    \Rho)
  \end{array}$}\\
// \com{$\rhobest$ is best generalization of $\rho$}\\
$\rhobest = \rho$\\
// \com{$\gen[i][j]$ is a generalization of $\rho$ using}\\
// \com{$\cc'[i]$}\\
$\gen = \mbox{new Rule[\cc.{\rm length}][3]}$\\
\forloop\ $i$ = 1  \forloopto\ $\cc$.length\\
~~  $f=\cc[i]$\\
~~     // \com{generalize by adding $f$ and eliminating}\\
~~     // \com{conjuncts for both attributes used in $f$.}\\
~~     $\gen[i][1] = \langle
\begin{array}[t]{@{}l@{}}
  \uae(\rho)[\uattr(f)\mapsto\top],\\
  \rae(\rho)[\rattr(f)\mapsto\top],\\
  \ops(\rho), \con(\rho)\union\set{f}\rangle
\end{array}$\\
~~ // \com{generalize by adding $f$ and eliminating}\\
~~ // \com{conjunct for user attribute used in $f$}\\
~~ $\gen[i][2] = \langle
\begin{array}[t]{@{}l@{}}
  \uae(\rho)[\uattr(f)\mapsto\top], \rae(\rho),\\
  \ops(\rho), \con(\rho)\union\set{f}\rangle
\end{array}$\\
~~ // \com{generalize by adding $f$ and eliminating}\\
~~ // \com{conjunct for resource attrib.\ used in $f$.}\\
~~ $\gen[i][3] = \langle
\begin{array}[t]{@{}l@{}}
  \uae(\rho), \rae(\rho)[\rattr(f)\mapsto\top],\\
  \ops(\rho), \con(\rho)\union\set{f}\rangle
\end{array}$\\
\eforloop\\
\forloop\ $i$ = 1 \forloopto\ $\cc$.length {\bf and} $j$ = 1 \forloopto\ 3\\
~~ // \com{try to further generalize $\gen[i]$}\\
~~ $\rho'' = \generalizeruleL(
\begin{array}[t]{@{}l@{}}
  \gen[i][j], \cc[i\!+\!1\,..],\\
  \uncovup, \Rho)
\end{array}$\\
~~ \ifstmt\ $\QrulL(\rho'', \uncovup) > \QrulL(\begin{array}[t]{@{}l@{}}
  \rhobest, \\
  \uncovup)
\end{array}$\\
~~ ~~ $\rhobest = \rho''$\\
~~\eifstmt\\
\eforloop\\
\return\ $\rhobest$
\end{tabular}
\begin{tabular}[t]{@{}l@{}}
\hypertarget{mergerules}{\function\ \mergerules(\Rho)}\\
// \com{Remove redundant rules}\\
$\redrules = 
\begin{array}[t]{@{}l@{}}
  \{\rho\in\Rho \;|\; \exists\, \rho' \in \Rho\setminus\set{\rho}.\\
  \hspace*{2em} \mean{\rho}\intersect\up_0\subseteq\mean{\rho'}\intersect\up_0\}
\end{array}$\\
$\Rho.\removeall(\redrules)$\\
// \com{Merge rules}\\
$\wkset = \{
\begin{array}[t]{@{}l@{}}
 (\rho_1,\rho_2) \;|\; \rho_1 \in \Rho \land \rho_2 \in \Rho\\
  \hspace*{1em} {} \land \rho_1\ne \rho_2 \land \con(\rho_1)=\con(\rho_2)\}  
\end{array}$\\
\whileloop\ not(\wkset.empty())\\
~~     $(\rho_1,\rho_2) = \wkset.{\rm remove}()$\\
~~     $\rhomerge = \langle
\begin{array}[t]{@{}l@{}}
  \uae(\rho_1)\union\uae(\rho_2),\\
  \rae(\rho_1)\union\rae(\rho_2),\\
  \ops(\rho_1)\union\ops(\rho_2), \con(\rho_1)\rangle  
\end{array}$\\
~~ // \com{Find rules that become redundant}\\
~~ // \com{if merged rule $\rhomerge$ is added}\\
~~ $\redrules = \setc{\rho\in\Rho}{\mean{\rho}\subseteq\mean{\rhomerge}}$\\
~~ // \com{Add the merged rule and remove redun-}\\
~~ // \com{dant rules if this improves policy quality}\\
~~ // \com{and the merged rule does not have}\\
~~ // \com{over-assignments.}\\
~~ \ifstmt\ $\begin{array}[t]{@{}l@{}}
  (\Qpol(\Rho \union \{\rhomerge\} \setminus \redrules) < \Qpol(\Rho)\\
  ~~ {} \land \mean{\rhomerge}\subseteq\up_0)
\end{array}$\\
~~ ~~ $\Rho.\removeall(\redrules)$\\
~~ ~~ $\wkset.\removeall(\{(\rho_1,\rho_2) \in \wkset \;|\; $\\
~~ ~~ \hspace*{2em} ${} \rho_1 \in \redrules \vee \rho_2 \in \redrules\})$\\
~~ ~~ $\wkset.\addall(
\begin{array}[t]{@{}l@{}}
  \{(\rhomerge,\rho) \;|\; \rho \in \Rho\\
  {} \land \con(\rho)=\con(\rhomerge)\})  
\end{array}$\\
~~ ~~ $\Rho.\add(\rhomerge)$\\
~~ \eifstmt\\
\ewhileloop\\
\return\ $\true$ if any rules were merged\\
\end{tabular}
\caption{Left: Generalize rule $\rho$ by adding some formulas from $\cc$ to
  its constraint and eliminating conjuncts for attributes used in those
  formulas.  $f[x \mapsto y]$ denotes a copy of function $f$ modified so
  that $f(x)=y$.  $a[i..]$ denotes the suffix of array $a$ starting at
  index $i$. Right: Merge pairs of rules in $\Rho$, when possible, to
  reduce the WSC of $\Rho$.  $(a,b)$ denotes an unordered pair with
  components $a$ and $b$.  The union $e=e_1 \union e_2$ of attribute
  expressions $e_1$ and $e_2$ over the same set $A$ of attributes is
  defined by: for all attributes $a$ in $A$, if $e_1(a)=\top$ or
  $e_2(a)=\top$ then $e(a)=\top$ otherwise $e(a)=e_1(a)\union e_2(a)$.
}
\label{fig:generalizerule}
\label{fig:mergerules}
\end{figure}

\fullonly{
\section{Functions to Simplify Rules}
\label{sec:simplify-rules}

The function $\simplifyrulesL(\Rho)$ in Figure \ref{fig:simplifyrules}
attempts to simplify all of the rules in $\Rho$.  It updates its argument
$\Rho$ in place, replacing rules in $\Rho$ with simplified versions when
simplification succeeds.  It returns a Boolean indicating whether any rules
were simplified.  It attempts to simplify each rule in several ways, which
are embodied in the following functions that it calls.  The names of these
functions start with ``elim'', because they attempt to eliminate
unnecessary parts of rules. To enable $\simplifyrulesL$ to determine
whether any rules were simplified, each ``elim'' function returns a Boolean
value indicating whether it simplified any rules. For brevity, computation
of the Boolean return values of ``elim'' functions are not reflected in the
pseudocode.

The function $\elimredundantsetsL$ is described above.  It returns
$\false$, even if some redundant sets were eliminated, because elimination
of redundant sets does not affect the meaning or mergeability of rules, so
it need not trigger another iteration of merging and simplification.

The function $\elimconjunctsL(\rho, \Rho, \up)$ in Figure
\ref{fig:simplifyrules} attempts to increase the quality of rule $\rho$ by
eliminating some\fullonly{ of the} conjuncts.  It calls the function
$\elimconjunctshelperL(\rho, A, \up)$ in Figure \ref{fig:simplifyrules},
which considers all rules that differ from $\rho$ by
mapping a subset $A'$ of the tagged attributes in $A$ to $\top$ instead of
to a set of values; among the resulting rules that are valid, it returns
one with the highest quality.  A {\em tagged attribute} is a pair of the
form $\tuple{\stringlit{user}, a}$ with $a \in \Au$ or
$\tuple{\stringlit{res}, a}$ with $a \in \Ar$.  The set $\Aunrm$ in 
$\elimconjunctsL$ is a set of {\em unremovable} tagged attributes; it is a
parameter of the algorithm, specifying attributes that should not be
eliminated, because eliminating them increases the risk of generating an
overly general policy, i.e., a policy that might grant inappropriate
permissions when new users or new resources (hence new permissions) are
added to the system.  We use a combinatorial algorithm for
$\elimconjunctsL$ that evaluates all combinations of conjuncts that can be
eliminated, because elimination of one conjunct might prevent elimination
of another conjunct.  This algorithm makes $\elimconjunctsL$ worst-case
exponential in the numbers of user attributes and resource attributes that
can be eliminated while preserving validity of the rule; in practice the
number of such attributes is small.  $\elimconjunctsL$ also considers
whether to remove conjuncts from the user attribute expression or the
resource attribute expression first, because elimination of a conjunct in
one attribute expression might prevent elimination of a conjunct in the
other.  The algorithm could simply try both orders, but instead it uses a
heuristic that, in our experiments, is faster and almost as effective: if
$\maxConjunctSz(\eu) \ge \maxConjunctSz(\er)$ then eliminate conjuncts from
the user attribute expression first, otherwise eliminate conjuncts from the
resource attribute expression first, where
\hypertarget{maxConjunctSz}{$\maxConjunctSz(e)$} is the size (WSC) of the
largest conjunct in attribute expression $e$.
$\elimconjunctshelper$ calls the function
\hypertarget{elimattrib}{$\elimattrib(\rho,\ta)$}, which returns a copy of
rule $\rho$ with the conjunct for tagged attribute $\ta$ removed from the
user attribute expression or resource attribute expression as appropriate
(in other words, the specified attribute is mapped to $\bot$); pseudocode
for $\elimattrib$ is straightforward and omitted.

The function
\hypertarget{elimconstraints}{$\elimconstraints(\rho,\Rho,\up)$}
attempts to improve the quality of $\rho$ by removing unnecessary atomic
constraints from $\rho$'s constraint.  An atomic constraint is {\it
  unnecessary} in a rule $\rho$ if removing it from $\rho$'s constraint
leaves $\rho$ valid. Pseudocode for $\elimconstraintsL$ is analogous to
$\elimconjunctsL$, except it considers removing atomic constraints instead
of conjuncts from rules.


The function \hypertarget{elimelements}{$\elimelements(\rho)$} attempts to
decrease the WSC of rule $\rho$ by removing elements from sets in conjuncts
for multi-valued user attributes, if removal of those elements preserves
validity of $\rho$; note that, because $\subseteq$ is used in the semantics
of user attribute expressions, the set of user-permission tuples that
satisfy a rule is never decreased by such removals.  It would be reasonable
to use a combinatorial algorithm for $\elimelementsL$, in the same style as
$\elimconjunctsL$ and $\elimconstraintsL$, because elimination of one set
element can prevent elimination of another.  We decided to use a simple
linear algorithm for this function, for simplicity and because it is likely
to give the same results, because $\elimelementsL$ usually eliminates only
0 or 1 set elements per rule in our experiments.
Pseudocode for $\elimelementsL$ is straightforward and omitted.


The function \hypertarget{elimoverlapVal}{$\elimoverlapVal(\rho, \Rho)$}
attempts to decrease the WSC of rule $\rho$ by removing values from
conjuncts of attribute expressions in $\rho$ if there are other rules that
cover the affected user-permission tuples and have higher quality.
Specifically, a value $v$ in the conjunct for a user attribute $a$ in
$\rho$ is removed if there is another rule $\rho'$ in $\Rho$ such that (1)
$\attrib(\uae(\rho')) \subseteq \attrib(\uae(\rho))$ and
$\attrib(\rae(\rho')) \subseteq \attrib(\rae(\rho))$, (2) the conjunct of
$\uae(\rho')$ for $a$ contains $v$, (3) each conjunct of $\uae(\rho')$ or
$\rae(\rho')$ other than the conjunct for $a$ is either $\top$ or a
superset of the corresponding conjunct of $\rho$, and (4) $\con(\rho')
\subseteq \con(\rho)$.
The condition for removal of a value in the conjunct for a resource
attribute is analogous.  If a conjunct of $\uae(\rho)$ or $\rae(\rho)$
becomes empty, $\rho$ is removed from $\Rho$.  $\elimoverlapValL(\rho,
\Rho)$ returns true if it modifies or removes $\rho$, otherwise it returns
false.  Pseudocode for $\elimoverlapValL$ is straightforward and omitted.

The function \hypertarget{elimoverlapOp}{$\elimoverlapOp(\rho, \Rho)$}
attempts to decrease the WSC of rule $\rho$ by removing operations from
$\ops(\rho)$, if there are other rules that cover the affected
user-permission tuples.  Specifically, an operation $o$ is removed from
$\ops(\rho)$ if there is another rule $\rho'$ in $\Rho$ such that (1)
$\attrib(\uae(\rho')) \subseteq \attrib(\uae(\rho))$ and
$\attrib(\rae(\rho')) \subseteq \attrib(\rae(\rho))$, (2) $\ops(\rho')$
contains $o$, (3) each conjunct of $\uae(\rho')$ or $\rae(\rho')$ is either
$\top$ or a superset of the corresponding conjunct of $\rho$, and (4)
$\con(\rho')$ is a subset of $\con(\rho)$.  If $\ops(\rho)$ becomes empty,
$\rho$ is removed from $\Rho$.  $\elimoverlapOpL(\rho, \Rho)$ returns true
if it modifies or removes $\rho$, otherwise it returns false.  Pseudocode
for $\elimoverlapOpL$ is straightforward and omitted.

\begin{figure}[tbp]
\articleonly{\begin{spacing}{1.0}}
\begin{tabbing}
\hypertarget{simplifyrules}{\function\ $\simplifyrules(\Rho)$}\\
\forloop\ $\rho$ \forloopin\ $\Rho$\\
~~\= $\elimredundantsetsL(\uae(\rho))$\\
\> $\elimconjunctsL(\rho,\Rho, \up_0)$\\
\> $\elimelementsL(\rho)$\\
\eforloop\\
\forloop\ $\rho$ \forloopin\ $\Rho$\\
\> $\elimoverlapValL(\rho,\Rho)$\\
\eforloop\\
\forloop\ $\rho$ \forloopin\ $\Rho$\\
\> $\elimoverlapOpL(\rho,\Rho)$\\
\eforloop\\
\forloop\ $\rho$ \forloopin\ $\Rho$\\
\> $\elimconstraintsL(\rho, \Rho, \up_0)$\\
\eforloop\\
\return\ $\true$ if any $\rho$ in $\Rho$ was changed\\
\\
\hypertarget{elimconjuncts}{\function\ $\elimconjuncts(\rho,\Rho, \up)$}\\
$Au = \set{\stringlit{user}} \times \attrib(\uae(\rho)) \setminus \Aunrm$\\
$Ar = \set{\stringlit{res}} \times \attrib(\rae(\rho)) \setminus \Aunrm$\\
\ifstmt\ $\maxConjunctSzL(\uae(\rho)) \ge \maxConjunctSzL(\rae(\rho))$\\
~~~~\= $\rho'=\elimconjunctshelperL(\rho, Au, \up)$\\
\>     $\rho''=\elimconjunctshelperL(\rho', Ar, \up)$\\
\elsestmt\\
\>     $\rho'=\elimconjunctshelperL(\rho, Ar, \up)$\\
\>     $\rho''=\elimconjunctshelperL(\rho', Au, \up)$\\
\eifstmt\\
\ifstmt\ $\rho'' \ne \rho$\\
~~~~\= replace $\rho$ with $\rho''$ in $\Rho$\\
\eifstmt\\
\\
\hypertarget{elimconjunctshelper}{\function\ $\elimconjunctshelper(\rho,A, \up)$}\\
$\rhobest = \rho$ \\
// \com{Discard tagged attributes $\ta$ such that elimi-}\\
// \com{nation of the conjunct for $\ta$ makes $\rho$ invalid.}\\
\forloop\ $\ta$ \forloopin\ $A$\\
~~~~\=  $\rho' = \elimattribL(\rho,\ta)$\\
\>     \ifstmt\ not $\mean{\rho'} \subseteq \up_0$\\
\>~~~~\=  $A.\removeElt(\ta)$\\
\>     \eifstmt\\
\eforloop\\
\forloop\ $i$ = 1 \forloopto\ $A$.length ~~~ // \com{treat $A$ as an array}\\
~~~~\= $\rho' = \elimattribL(\rho,A[i])$\\
~~~~\= $\rho'' = \elimconjunctshelperL(\rho', A[i\!+\!1\,..])$\\
\>     \ifstmt\ $\QrulL(\rho'', \up) > \QrulL(\rhobest, \up)$\\
\>~~~~\=  $\rhobest = \rho''$\\
\>     \eifstmt \\
\eforloop\\
\return\ $\rhobest$
\end{tabbing}
\articleonly{\end{spacing}}
\caption{Functions used to simplify rules. }
\label{fig:simplifyrules}
\end{figure}

}

\subsection{Example}
\label{sec:algorithm:running-example}


We illustrate the algorithm on a small fragment of our university case
study ({\it cf.} Section \ref{sec:evaluation:abac-policies}).  The fragment
contains a single rule $\rho_0=\tuple{\true, type \in \{{\rm
    gradebook}\},\linebreak[0] \{{\rm addScore}, {\rm readScore}\}, {\rm
    crsTaught} \ni {\rm crs}}$ and all of the attribute data from the full
case study, except attribute data for gradebooks for courses other than
cs601.  We consider an operation log $L$ containing three entries:
$\{\langle {\rm csFac2}, {\rm cs601gradebook},\lncsonly{\linebreak[0]} {\rm
  addScore}, {\rm t_1}\rangle, \langle {\rm
  csFac2},\articleonly{\linebreak[0]} {\rm cs601gradebook}, {\rm
  readScore}, {\rm t_2}\rangle,$ $\langle {\rm csStu3},$ ${\rm
  cs601gradebook},\lncsonly{\linebreak[4]}$ ${\rm addScore},$ ${\rm
  t_3}\rangle\}$.  User {\rm csFac2} is a faculty in the computer science
department who is teaching cs601; attributes are ${\rm position}={\rm
  faculty}$, ${\rm dept}={\rm cs}$, and ${\rm crsTaught}=\{{\rm cs601}\}$.
{\rm csStu3} is a CS student who is a TA of cs601; attributes are ${\rm
  position}={\rm student}$, ${\rm dept}={\rm cs}$, and ${\rm
  crsTaught}=\{{\rm cs601}\}$.  {\rm cs601gradebook} is a resource with
attributes ${\rm type}={\rm gradebook}$, ${\rm dept}={\rm cs}$, and ${\rm
  crs}={\rm cs601}$.

Our algorithm selects user-permission tuple $\tuple{{\rm csFac2}, {\rm
    cs601gradebook}, {\rm addScore}}$ as the first seed, and calls function
$\candidateconstraint$ to compute the set of atomic constraints that hold
between {\rm csFac2} and {\rm cs601gradebook}; the result is $\cc = \{{\rm
  dept} = \rm {dept}, {\rm crsTaught} \ni {\rm crs}\}$.
$\addcandidateruleL$ is called twice to compute candidate rules.  The first
call to $\addcandidateruleL$ calls $\computeuae$ to compute a UAE $\eu$
that characterizes the set $\su$ containing users with permission
$\tuple{{\rm addScore}, {\rm cs601gradebook}}$ and with the same candidate
constraint as {\rm csFac2} for {\rm cs601gradebook};
the result is $\eu=({\rm position} \in \{{\rm faculty, student}\} \land
{\rm dept}\in \{{\rm cs}\} \land {\rm crsTaught}\supseteq \{\{{\rm
  cs601}\}\})$.  $\addcandidateruleL$ also calls $\computerae$ to compute
a resource-attribute expression that characterizes $\set{{\rm
    cs601gradebook}}$; the result is $\er = ({\rm crs} \in \{{\rm cs601}\}
\land {\rm dept}\in \{{\rm cs}\} \land {\rm type} \in \{{\rm
  gradebook}\})$.  The set of operations considered in this call to
$\addcandidateruleL$ is simply $\so = \set{{\rm addScore}}$.
$\addcandidateruleL$ then calls $\generalizerule$, which generates a
candidate rule $\rho_1$ which initially has $\eu$, $\er$ and $\so$ in the
first three components, and then atomic constraints in $\cc$ are added to
$\rho_1$, and conjuncts in $\eu$ and $\er$ for attributes used in $\cc$ are
eliminated; the result is $\rho_1=\langle {\rm position \in \{faculty,
  student\}, type \in \{gradebook\},}$ ${\rm \{addScore\}, dept=dept \land
  crsTaught \ni crs}\rangle$, which also covers\fullonly{ the entitlement
  in} the third log entry.
Similarly, the second call to $\addcandidateruleL$ generates a candidate
rule $\rho_2 = \langle {\rm position \in \{faculty\}, type \in
  \{gradebook\},}$ ${\rm \{addScore, readScore\},}$ ${\rm dept=dept
  \land{}}$ ${\rm crsTaught \ni crs}\rangle$, which also covers\fullonly{
  the entitlement in} the second log entry.

All of $\up(L)$ is covered, so our algorithm calls $\mergerulesL$, which
attempts to merge $\rho_1$ and $\rho_2$ into rule $\rho_3=\langle {\rm
  position \in \{faculty, student\}, type \in \{gradebook\},}$ ${\rm
  \{addScore,}$ ${\rm readScore\}, dept=dept \land crsTaught \ni
  crs}\rangle$.
$\rho_3$ is discarded because it introduces an over-assignment\fullonly{
  user-permission tuple 
$\langle {\rm csStu3}, {\rm cs601gradebook},$ ${\rm readScore} \rangle$} 
while $\rho_1$ and $\rho_2$ do not.  Next,
$\simplifyrulesL$ is called, which first simplifies $\rho_1$ and $\rho_2$
to $\rho'_1$ and $\rho'_2$, respectively, and then eliminates $\rho'_1$
because it covers a subset of the tuples covered by $\rho'_2$.  The final
result is $\rho'_2$, which is identical to the rule $\rho_0$ in the
original policy.


\noise{
\subsection{Noise Detection}

If the input might contain noise, in the form of log entries representing
uses of permissions that should not be granted, then the algorithm should
allow the generated policy to have under-assignments (i.e., it should allow
$\up(L)\setminus\mean{\pi}$ to be non-empty), to reduce the amount of such
noise that gets propagated into the policy.  Such noise in the logs is due
to over-assignments in the currently deployed access control policy.  We
adopt the approach in \cite{xu13miningABAC} to identify suspected
over-assignment noise.  The main idea is that rules with quality below a
threshold $\tau$ are considered to represent noise and hence are omitted
from the generated policy, and that user-permission tuples in $\up(L)$ that
are represented only by rules classified as noise are reported as suspected
over-assignment noise.  An additional consideration in this context is that
usage frequency should be considered: under-assignments in the generated
policy should correspond to infrequently used entitlements.  Thus, the rule
quality metric used to identify rules that represent noise should be
frequency-sensitive.  One option is to use the frequency-sensitive rule
quality metric $\fwQrulL$.  Another option is to use a metric focused
exclusively on the usage frequency of the permissions covered by a
rule,\shortonly{ such as}\fullonly{ ignoring the size of the rule, the
  number of covered user-permission tuples, and the over-assignments (if
  any).  Such a metric is}
$Q_{\rm freq}(\rho, L) = |\mean{\rho}|^{-1} \sum_{t \in \mean{\rho}} \freqL(t, L)$.


The rule quality metric used for noise detection can be the same rule
quality metric used in the main part of the algorithm, in which case the
main part of the algorithm is modified to stop adding candidate rules to
the final policy when the rule quality drops below threshold, or a
different metric, in which case an additional loop is added at the end of
the algorithm to find and remove rules whose quality is below threshold.
}


\section{Evaluation Methodology}
\label{sec:evaluation}

We evaluate our policy mining algorithms on synthetic operation logs
generated from ABAC policies (some handwritten and some synthetic) and
probability distributions characterizing the frequency of actions.  This
allows us to evaluate the effectiveness of our algorithm by comparing the
mined policies with the original ABAC policies.  We are eager to also
evaluate our algorithm on actual operation logs and actual attribute data,
when we are able to obtain them.


\subsection{ABAC Policies}
\label{sec:evaluation:abac-policies}

\myparagraph{Case Studies}

We developed four case studies for use in evaluation of our algorithm,
described briefly here.  Details of the case studies, including all policy
rules, various size metrics (number of users, number of resources, etc.),
and some illustrative attribute data, appear in \cite{xu13miningABAC}.

Our {\em university case study} is a policy that controls access by
students, instructors, teaching assistants, registrar officers, department
chairs, and admissions officers to applications (for admission),
gradebooks, transcripts, and course schedules.  Our {\em health care case
  study} is a policy that controls access by nurses, doctors, patients, and
agents (e.g., a patient's spouse) to electronic health records (HRs) and HR
items (i.e., entries in health records).  Our {\em project management case
  study} is a policy that controls access by department managers, project
leaders, employees, contractors, auditors, accountants, and planners to
budgets, schedules, and tasks associated with projects.  Our {\em online
  video case study} is a policy that controls access to videos by users of
an online video service.

The number of rules in the case studies is relatively small ($10\pm 1$ for
the first three case studies, and 6 for online video), but they express
non-trivial policies and exercise all the features of our policy language,
including use of set membership and superset relations in attribute
expressions and constraints.  The manually written attribute dataset for
each case study contains a small number of instances of each type of user
and resource.

For the first three case studies, we generated a series of synthetic
attribute datasets, parameterized by a number $N$, which is the number of
departments for the university and project management case studies, and the
number of wards for the health care case study. The generated attribute
data for users and resources associated with each department or ward are
similar to but more numerous than the attribute data in the manually
written datasets.  We did not bother creating synthetic data for the online
video case study, because the rules are simpler.



\myparagraph{Synthetic Policies}

We generated synthetic policies using the algorithm proposed by Xu and
Stoller \cite{xu13miningABAC}.  Briefly, the policy synthesis algorithm
first generates the rules and then uses the rules to guide generation of
the attribute data; this allows control of the number of granted
permissions.  The algorithm takes $\Nrule$, the desired number of rules, as
an input. The numbers of users and resources are proportional to the number
of rules.  Generation of rules and attribute data is based on several
probability distributions, which are based loosely on the case studies or
assumed to have a simple functional form (e.g., uniform distribution).

\subsection{Log Generation}
\label{sec:log-gen}

\shortonly{
The inputs to the algorithm are an ABAC policy $\pi$, the desired
completeness of the log, and several probability distributions.  The {\em
  completeness} of a log, relative to an ABAC policy, is the fraction of
user-permission tuples in the meaning of the policy that appear in at least
one entry in the log.  A straightforward log generation algorithm would
generate each log entry by first selecting an ABAC rule, according to a
probability distribution on rules, and then selecting a user-permission
tuple that satisfies the rule, according to probability distributions on
users, resources, and operations.  This process would be repeated until the
specified completeness is reached.  This algorithm is inefficient when high
completeness is desired.  Therefore, we adopt a different approach that
takes advantage of the fact that our policy mining algorithm is insensitive
to the order of log entries and depends only on the frequency of each
user-permission tuple in the log.  In particular, instead of generating
logs (which would contain many entries for popular user-permission tuples),
our algorithm directly generates a {\em log summary}, which is a set of
user-permission tuples with associated frequencies (equivalently, a set of
user-permission tuples and a frequency function).

}\fullonly{
We first describe a straightforward but inefficient log generation
algorithm, and then describe the log summary generation algorithm that is
actually used in our experiments.

\myparagraph{Log Generation Algorithm}

The inputs to the algorithm are an ABAC policy $\pi$, the desired
completeness of the log, and several probability distributions.  The {\em
  completeness} of a log, relative to an ABAC policy, is the fraction of
user-permission tuples in the meaning of the policy that appear in at least
one entry in the log.  We use completeness as a measure of log size
relative to policy size.  The algorithm generates each log entry by first
selecting an ABAC rule, according to a probability distribution on roles,
and then selecting a user-permission tuple that satisfies the rule,
according to probability distributions on users, resources, and operations.
This process is repeated until the specified completeness is reached.

Specifically, our log generation algorithm uses the following probability
distributions.  For each rule $\rho$ in $\pi$, $\Prule(\rho)$ is the
probability of selecting $\rho$.  For each operation $o$ and rule $\rho$,
$\Pop(o|\rho)$ is the conditional probability of selecting operation $o$
when $\rho$ has been selected; for sanity, we require $\Pop(o|\rho) = 0$ if
$o\not\in\ops(\rho)$.  For each user $u$, resource $r$, and rule $\rho$,
$\PuserRes(u,r|\rho)$ is the conditional probability of selecting user $u$
and resource $r$ when rule $\rho$ and resource $r$ have been selected; for
sanity, we require $\PuserRes(u,r|\rho)=0$ if $u\not\models\uae(\rho) \lor
r\not\models\rae(\rho) \lor \tuple{u,r}\not\models\con(\rho)$.  The
algorithm generates each log entry as follows: select a rule $\rho$ based
on $\Prule(\rho)$, select an operation $o$ based on $\Pop(o|\rho)$ and a
user-permission tuple $\tuple{u,r}$ based on $\PuserRes(u,r|\rho)$, and
then create a log entry $\tuple{u,r,o,i}$, where the unique identifier $i$
is the index of this log entry in the synthetic log.

For convenience, we allow the user to specify probability distributions
$\Puser(u)$, $\Pres(r)$, and $\PopSimp(o)$ instead of $\PuserRes$ and
$\Pop$, and we compute $\PuserRes$ and $\Pop$ from them as described below;
intuitively, the probability associated with invalid values is distributed
uniformly among the valid values by this computation.  These alternative
probability distributions provide less detailed control but are simpler and
more convenient.
\begin{eqnarray*}
  \label{eq:loggen-dist}
  \PuserRes(u,r|\rho) &=&
  \left\{\begin{array}{@{}ll@{}}
      c^{-1}\Puser(u)\Pres(r) & {\rm if}~\tuple{u,r}\in \mean{\rho}_{\rm ur}\\
      0 & {\rm otherwise}
    \end{array}\right.\\
  && \mbox{where } c=\sum_{\tuple{u,r} \in \mean{\rho}_{\rm ur}}\Puser(u)\Pres(r)\\
  \Pop(o|\rho) &=&
  \left\{\begin{array}{@{}ll@{}}
      c^{-1}\PopSimp(o) & {\rm if}~o \in \ops(\rho)\\
      0 & {\rm otherwise}
    \end{array}\right.\\
  && \mbox{where } c=\sum_{o \in \ops(\rho)}\PopSimp(o)
\end{eqnarray*}
where the user-resource part of the meaning of a rule is $\mean{\rho}_{\rm
  ur}=\UNION_{\tuple{u,r,o}\in\mean{\rho}} \{\tuple{u,r}\}$.

\myparagraph{Log Summary Generation Algorithm}

When the probability distributions give much higher probability to some
user-permission tuples than others, using the above algorithm to generate
logs with high completeness is very inefficient, because very long logs,
with many entries for high-probability user-permission tuples, are
generated.  Therefore, we adopt a different approach that takes advantage
of the fact that our policy mining algorithm is insensitive to the order of
log entries and depends only on the frequency of each user-permission tuple
in the log.  Specifically, instead of generating logs, we directly generate
{\em log summaries}, which are simply sets of user-permission tuples with
associated frequencies that sum to 1 (in other words, a log summary defines
a set of user-permission tuples and a frequency function).  For a given
ABAC policy and probability distributions, we generate a log summary with
completeness 1 using a program that computes the asymptotic frequency with
which each tuple in $\up_0$ would appear in an infinitely long log
generated using the above algorithm.
To generate a series of log summaries with varying completeness, we start
with the log summary with completeness 1 and repeatedly apply the following
procedure that generates a log summary of lower completeness from a log
summary with higher completeness: select the appropriate number of
user-permission tuples from the latter, with the selection probability for
each tuple equal to its associated frequency, and then normalize the
frequencies of the selected tuples so they sum to 1.
}

\myparagraph{Probability Distributions}

An important characteristic of the probability distributions used in
synthetic log and log summary generation is the ratio between the most
frequent (i.e., most likely) and least frequent items of each type (rule,
user, etc.).  For case studies with manually written attribute data, we
manually created probability distributions in which this ratio ranges from
about 3 to 6.  For case studies with synthetic data and synthetic policies,
we generated probability distributions in which this ratio is 25 for rules,
25 for resources, 3 for users, and 3 for operations (the ratio for
operations has little impact, because it is relevant only when multiple
operations appear in the same rule, which is uncommon).

\subsection{Metrics}
\label{sec:evaluation:methodology}

For each case study and each associated attribute dataset (manually written
or synthetic), we generate a synthetic operation log using the algorithm in
Section \ref{sec:log-gen} and then run our ABAC policy mining algorithms.
We evaluate the effectiveness of each algorithm by comparing the generated
ABAC policy to the original ABAC policy, using the metrics described below.

\myparagraph{Syntactic Similarity}


Jaccard similarity of sets is $J(S_1, S_2)=|S_1\intersect S_2| \,/\, |S_1
\union S_2|$.  Syntactic similarity of UAEs is defined by $\uaeSim(e,e') =
|\Au|^{-1}\sum_{a\in\Au} J(e(a), e'(a))$.  Syntactic similarity of RAEs is
defined by $\raeSim(e,e') = |\Ar|^{-1}\sum_{a\in\Ar} J(e(a), e'(a))$.  The
syntactic similarity of rules $\tuple{\eu, \er, O, c}$ and
$\tuple{\eu',\er',O',c'}$ is the average of the similarities of their
components, specifically, the average of $\uaeSim(\eu, \eu')$,
$\raeSim(\er, \er')$, $J(O, O')$, and $J(c, c')$.  The {\em syntactic
  similarity} of rule sets $\Rho$ and $\Rho'$ is the average, over rules
$\rho$ in $\Rho$, of the syntactic similarity between $\rho$ and the most
similar rule in $\Rho'$.  The {\em syntactic similarity} of policies $\pi$
and $\pi'$ is the maximum of the syntactic similarities of the sets of
rules in the policies, considered in both orders (this makes the relation
symmetric).\fullonly{ Formally,
  \begin{eqnarray*}
    \synSim(\Rho,\Rho') &=& |\Rho|^{-1}
    \sum_{\rho\in\Rho}\min(\setc{\synSim(\rho,\rho')}{\rho'\in \Rho'}))\\
    \synSim(\pi,\pi') &=& \max(\synSim(\rules(\pi), \rules(\pi')),
    \synSim(\rules(\pi'), \rules(\pi)))
  \end{eqnarray*}}  
Syntactic similarity ranges from 0 (completely different) to 1
(identical).\fullonly{ High syntactic similarity between the
  original and generated ABAC policies is not always desirable, e.g., for
  synthetic policies, which tend to have unnecessarily complicated rules.}
%


\myparagraph{Semantic Similarity}

Semantic similarity measures the similarity of the entitlements granted by
two policies.  The {\em semantic similarity} of policies $\pi$ and $\pi'$
is defined by $J(\mean{\pi}, \mean{\pi'})$.  Semantic similarity ranges
from 0 (completely different) to 1 (identical).

\myparagraph{Fractions of Under-Assignments and Over-Assignments}

To characterize the semantic differences between an original ABAC policy
$\pi_0$ and a mined policy $\pi$ in a way that distinguishes
under-assignments and over-assignments, we compute the fraction of
over-assignments and the fraction of under-assignments, defined by
$|\mean{\pi}\setminus\mean{\pi_0}|\,/\,|\mean{\pi}|$ and
$|\mean{\pi_0}\setminus\mean{\pi}|\,/\,|\mean{\pi}|$, respectively.

\noise{
\subsection{Noise}
\label{sec:evaluation:noise}

We consider two kinds of noise: permission noise in form of
over-assignments (i.e., incorrectly granted permissions), and attribute
noise.  We do not consider under-assignments (i.e., absence of permissions
that should be granted), because introducing under-assignments is
equivalent to decreasing the log completeness.

To evaluate the effectiveness of our noise detection techniques in the
presence of over-assignment noise with level $\nu_o$ and attribute noise
with level $\nu_a$, we start with an ABAC policy, change attribute values
until the validity of $\nu_a|\up_0|$ user-permission tuples is changed,
generate a synthetic log summary from the given ABAC rules and the modified
attribute data, with the log summary generation algorithm modified to
reflect that each log entry has a probability $\nu_o$ of embodying an
over-assignment instead of a valid user-permission tuple, and then apply
our policy mining algorithm to the resulting log summary and the modified
attribute data.  The over-assignments are invalid user-permission tuples
generated according to the same probability distributions used in synthetic
log generation.  The changed attribute values are divided equally between
missing values (i.e., replace a non-bottom value with bottom) and incorrect
values (i.e., replace a non-bottom value with another non-bottom value).

Our current noise detection technique does not attempt to distinguish
over-assignment noise from attribute noise (this is a topic for future
research); policy analysts are responsible for determining whether a
reported suspected error is due to an incorrect permission, an incorrect or
missing attribute value, or a false alarm.  Therefore, when comparing
actual over-assignments to reported suspected over-assignments, permission
changes due to attribute noise (i.e., changes in the set of user-permission
tuples that satisfy the policy rules) are included in the actual noise.
}


\section{Experimental Results}
\label{sec:experiments}

This section presents experimental results using an implementation of our
algorithm in Java.  The implementation, case studies, and synthetic
policies used in the experiments are available at
\url{http://www.cs.stonybrook.edu/~stoller/}.

\myparagraph{Over-Assignment Weight}

The optimal choice for the over-assignment weights $w_o$ and $w'_o$ in the
policy quality and rule quality metrics, respectively, depends on the log
completeness.  When log completeness is higher, fewer over-assignments are
desired, and larger over-assignments weights give better results.  In
experiments, we take $w_o = 50 c - 15$ and $w'_o=w_o/10$, where $c$ is log
completeness.  In a production setting, the exact log completeness would be
unknown, but a rough estimate suffices, because our algorithm's results are
robust to error in this estimate.  For example, for case studies with
manually written attribute data, when the actual log completeness is 80\%,
and the estimated completeness used to compute $w_o$ varies from 70\% to
90\%, the semantic similarity of the original and mined policies varies by
0.04, 0.02, and 0 for university, healthcare, and project management,
respectively.


\myparagraph{Experimental Results}

Figure \ref{fig:results} shows results from our algorithm.  In each graph,
curves are shown for the university, healthcare, and project management
case studies with synthetic attribute data with $N$ equal to 6, $10$, and
$10$, respectively (average over results for 10 synthetic datasets, with 1
synthetic log per synthetic dataset), the online video case study with
manually written attribute data (average over results for 10 synthetic
logs), and synthetic policies with $\Nrule=20$ (average over results for 10
synthetic policies, with 1 synthetic log per policy).  Error bars show
standard deviation. Running time is at most 12 sec for each problem
instance in our experiments.

For log completeness 100\%, all four case study policies are reconstructed
exactly\fullonly{ (i.e., syntactic similarity and semantic similarity are
  1)}, and the semantics of synthetic policies is reconstructed almost
exactly: the semantic similarity is 0.99.  This is a non-trivial result,
especially for the case studies: an algorithm could easily generate a
policy with over-assignments or more complex rules.  As expected, the
results get worse as log completeness decreases.  When evaluating the
results, it is important to consider what levels of log completeness are
likely to be encountered in practice.  One datapoint comes from Molloy {\it
  et al.}'s work on role mining from real logs
\shortonly{\cite{molloy12generativeShort}}\fullonly{\cite{molloy12generative}}.
For the experiments in \shortonly{\cite[Tables 4 and
  6]{molloy12generativeShort}}\fullonly{\cite[Tables 4 and
  6]{molloy12generative}}, the actual policy is not known, but their
algorithm produces policies with 0.52\% or fewer over-assignments relative
to $\up(L)$, and they interpret this as a good result, suggesting that they
consider the log completeness to be near 99\%.  Based on this, we consider
our experiments with log completeness below 90\% to be severe stress tests,
and results for log completeness 90\% and higher to be more representative
of typical results in practice.


Syntactic similarity for all four case studies is above 0.91 for log
completeness 60\% or higher, and is above 0.94 for log completeness 70\% or
higher.  Syntactic similarity is lower for synthetic policies, but this is
actually a good result.  The synthetic policies tend to be unnecessarily
complicated, and the mined policies are better in the sense that they have
lower WSC.  For example, for 100\% log completeness, the mined policies
have 0.99 semantic similarity to the synthetic policies (i.e., the meaning
is almost the same), but the mined policies are simpler, with WSC 17\% less
than the original synthetic policies.

Semantic similarity is above 0.85 for log completeness 60\% or higher, and
above 0.94 for log completeness 80\% or higher.
These results are quite good, in the sense that our algorithm compensates
for most of the log incompleteness.  For example, at log completeness 0.6,
for policies generated by a policy mining algorithm that produces policies
granting exactly the entitlements reflected in the log, the semantic
similarity would be 0.6.  With our algorithm, the semantic similarity,
averaged over the 5 examples, is 0.95.  Thus, in this case, our algorithm
compensates for 35/40 = 87.5\% of the incompleteness.

The fractions of over-assignments are below 0.03 for log completeness 60\%
or higher.  The fractions of under-assignments are below 0.05 for log
completeness 60\% or higher for the case studies and are below 0.05 for log
completeness 80\% or higher for synthetic policies.  The graphs also show
that the semantic differences are due more to under-assignments than
over-assignments; this is desirable from a security perspective.


\articleonly{\newcommand{\graphwidth}{82mm}}
\lncsonly{\newcommand{\graphwidth}{60mm}}
\begin{figure}[tb]
  \centering
  \includegraphics[width=\graphwidth]{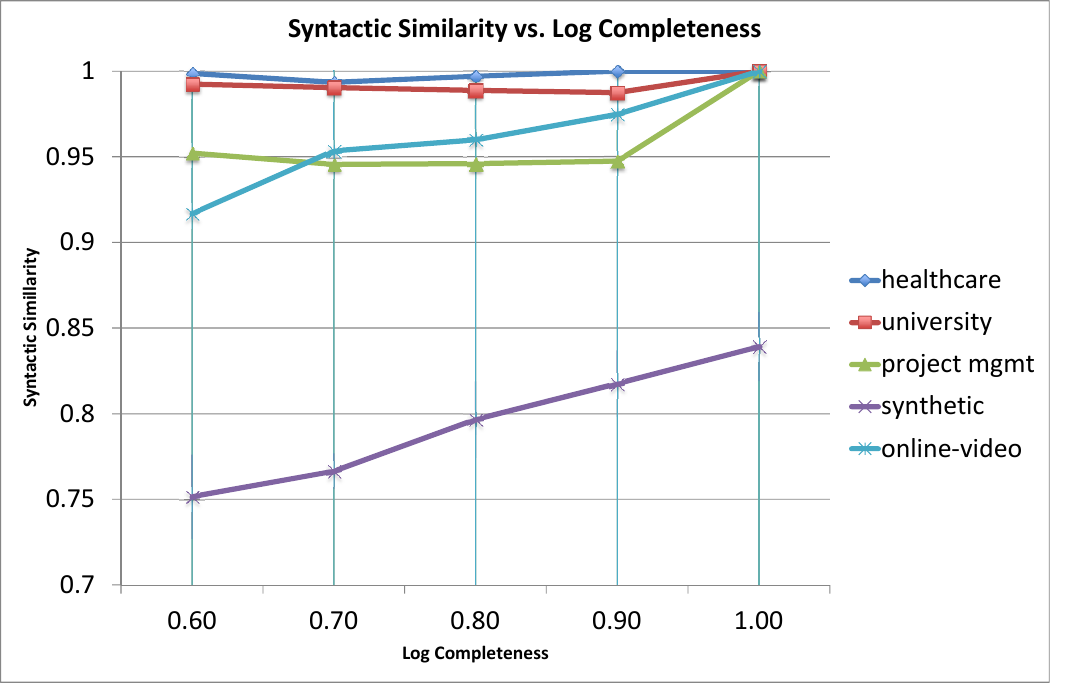}
  \includegraphics[width=\graphwidth]{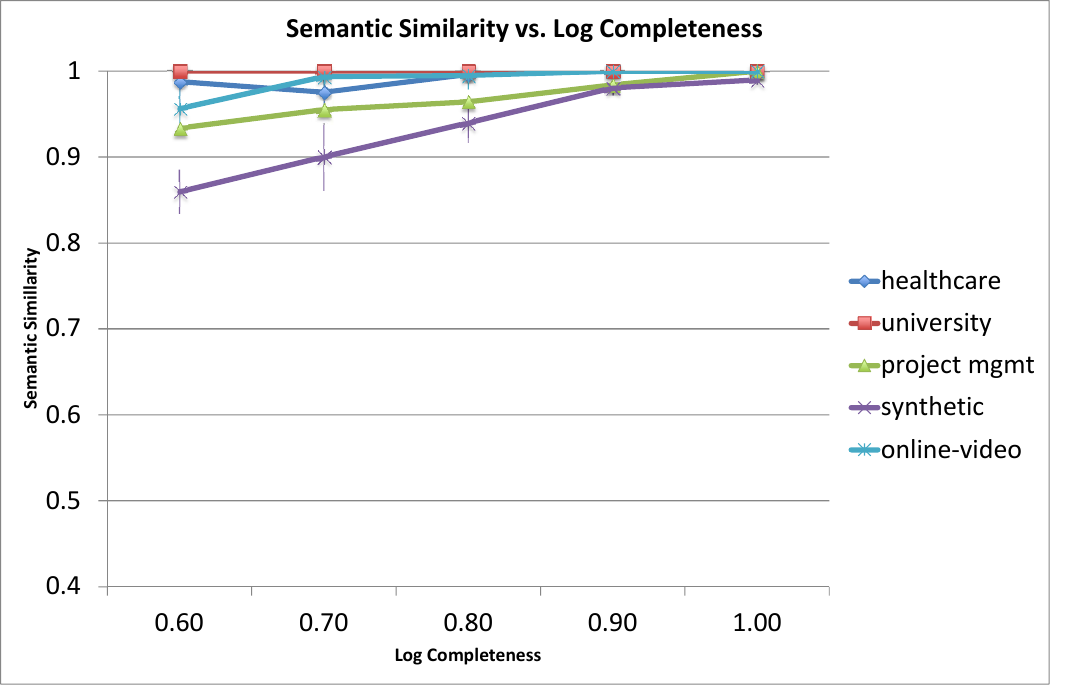}
  \includegraphics[width=\graphwidth]{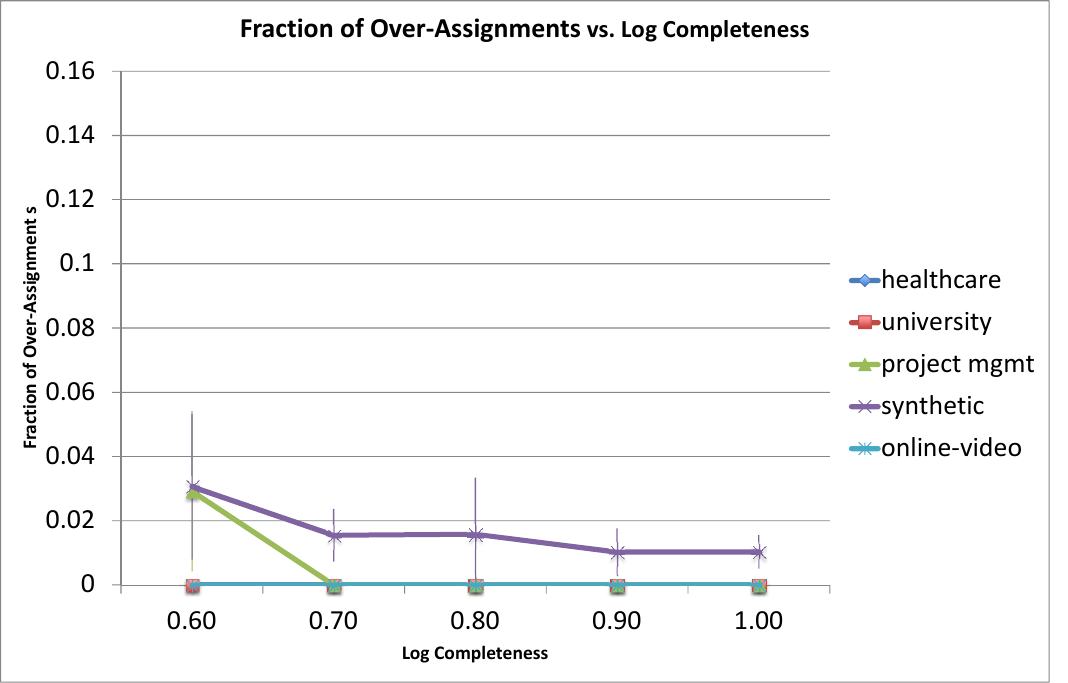}
  \includegraphics[width=\graphwidth]{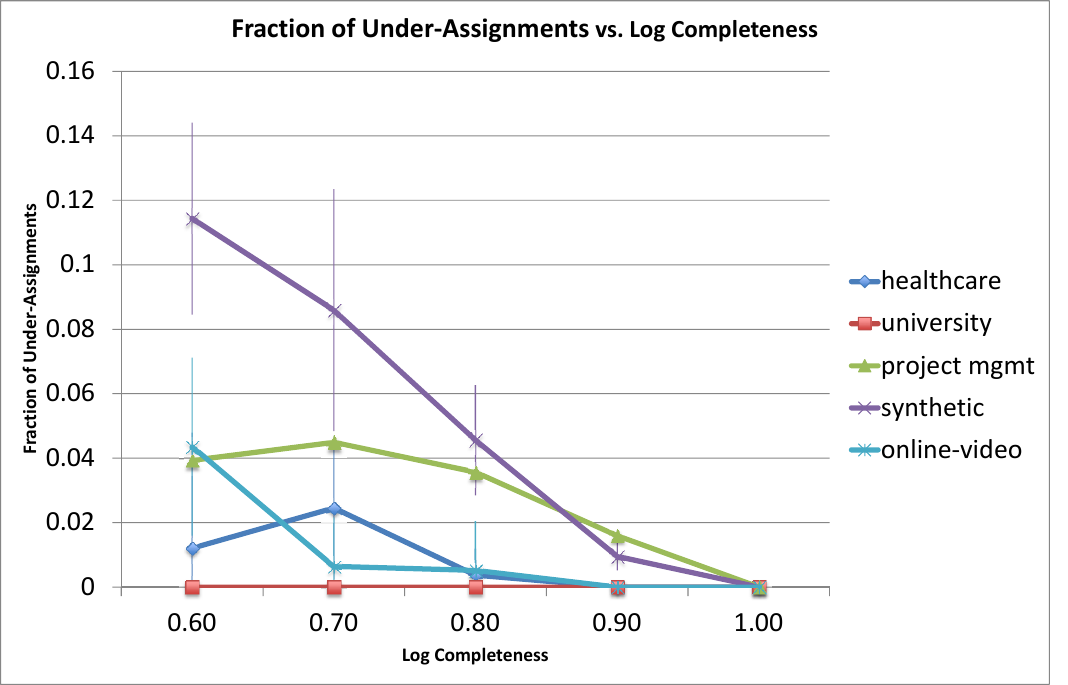}
  \caption{Top: Syntactic similarity and semantic similarity of original
    and mined ABAC policies, as a function of log completeness.  Bottom:
    Fractions of over-assignments and under-assignments in mined ABAC
    policy, as a function of log completeness. }
  \label{fig:results}
\end{figure}

 

\myparagraph{Comparison of Rule Quality Metrics}

The above experiments use the first rule quality metric, $\Qrul$, in
Section \ref{sec:algorithm}.  We also performed experiments using $\fwQrul$
and $\QrulILP$ on case studies with manually written attribute data and
synthetic policies.  $\Qrul$ is moderately better overall than $\fwQrul$
and significantly better overall than $\QrulILP$.

\myparagraph{Comparison with Inductive Logic Programming}
\label{sec:experiments:ilp}

To translate ABAC policy mining from logs to Progol
\shortonly{\cite{muggleton01cprogol}}\fullonly{\cite{muggleton96learning,muggleton01cprogol}},
we used the translation of ABAC policy mining from ACLs to Progol in
\cite[Sections 5.5, 16]{xu13miningABAC}, except negative examples
corresponding to absent user-permission tuples are omitted from the
generated program, and the statement {\tt set(posonly)?} is included,
telling Progol to use its algorithm for learning from positive examples.
For the four case studies with manually written attribute data (in
contrast, Figure \ref{fig:results} uses synthetic attribute for three of
the case studies), for log completeness 100\%, semantic similarity of the
original and Progol-mined policies ranges from 0.37 for project management
and healthcare to 0.93 for online video, while our algorithm exactly
reconstructs all four policies.\fullonly{ We used Progol 4.4 for these
  experiments, because Progol 5.0 segmentation-faults whenever the {\tt
    posonly} option is set.  We also ran the experiments without the {\tt
    posonly} option using Progol 4.4 and Progol 5.0, and we obtained the
  same results.}


% ...


\fullonly{
\myparagraph{Comparison with Author-Topic Model}
\label{sec:experiments:atm}


Inspired by Molloy {\it et al.}'s translation of RBAC policy mining from
logs to the problem of constructing an Author-Topic Model (ATM)
\cite{molloy12generative}, we developed and implemented a translation of
ABAC policy mining from logs to ATM.  In our translation, (1) topics
correspond to rules, (2) authors correspond to tuples containing a user
attribute expression, a resource attribute expression, and a constraint
that can appear together in a candidate ABAC rule, (3) documents correspond
to a user-resource pair, and (4) words correspond to operations.  We used
Steyvers and Griffiths' implementation of the ATM algorithm, available at
\url{http://psiexp.ss.uci.edu/research/programs_data/toolbox.htm}.  We did
not implement the discretization algorithm based on simulated annealing to
compute an ABAC policy from the generated author-topic model; we manually
considered the top few authors for each document.  We applied this
algorithm to the four case studies with manually written attribute data,
using generated logs with completeness ranging from 0.6 to 1 in steps of
0.1, and using 10 logs for each completeness level.  In each case, we
specified the number of rules in the original policy as the number of
topics.  In all cases (i.e., for all logs at each log completeness level
for each case study), the semantic similarity of the original and mined
policies is below 0.6.  The reason for the low semantic similarity is that
the authors selected by the ATM algorithm correspond to narrow rules, not
general rules applicable to multiple projects, departments, wards, or
whatever.  }


\section{Related Work}
\label{sec:related}

We are not aware of prior work on ABAC mining from logs.  We discuss prior
work on related problems.


Our policy mining algorithm is based on our algorithm for ABAC policy
mining from ACLs \cite{xu13miningABAC}.  The main differences are described
in Section \ref{sec:intro}.\fullonly{ The fact that our algorithm can be
  adapted to work well for this new problem, without changing the
  algorithm's overall structure, demonstrates the power and flexibility of
  our approach.}

Association rule mining is another possible basis for ABAC policy mining.
However, association rule mining algorithms are not well suited to ABAC
policy mining, because they are designed to find rules that are
probabilistic in nature\fullonly{ \cite{agrawal94fast}} and are supported
by statistically strong evidence.  They are not designed to produce a set
of rules 
that completely cover the input data and are minimum-sized among such sets
of rules.  Consequently, unlike our algorithm, they do not give preference
to smaller rules or rules with less overlap\fullonly{ (to reduce overall
  policy size)}.

Ni {\it et al.} investigated the use of machine learning algorithms for
security policy mining \cite{ni09automating}.
In the most closely related part of their work, a supervised machine
learning algorithm is used to learn classifiers (analogous to attribute
expressions) that associate users with roles, given as input the users, the
roles, user attribute data, and the user-role assignment.  Perhaps the
largest difference between their work and ABAC policy mining is that their
approach needs to be given the roles and the role-permission or user-role
assignment as training data; in contrast, ABAC policy mining algorithms do
not require any part of the desired high-level policy to be given as input.
Also, their work does not consider anything analogous to constraints.

Gal-Oz {\it et al.} \cite{gal-oz11miningShort} mine roles from logs that
record sets of permissions exercised together in one high-level operation.
Their algorithm introduces roles whose sets of assigned permissions are the
sets of permissions in the log.  Their algorithm introduces
over-assignments
by removing roles with few users or whose permission set occurs few times
in the log and re-assigning their members to roles with more
permissions.\fullonly{ They evaluate their algorithm on synthetic logs.}
Their algorithm does not use attribute data.

\shortonly{Molloy {\it et al.} apply a machine learning algorithm that uses
  a statistical approach, based upon a generative model, to find the RBAC
  policy that is most likely to generate the behavior (usage of
  permissions) observed in the logs \cite{molloy12generativeShort}.  They
  give an algorithm, based on Rosen-Zvi et al.'s algorithm for learning
  Author-Topic Models (ATMs), to mine meaningful roles from logs and
  attribute data, i.e., roles such that the user-role assignment is
  statistically correlated with user attributes.  Their approach can be
  adapted to ABAC policy mining from logs, but its scalability in this
  context is questionable, because the adapted algorithm would enumerate
  and then rank all tuples containing a UAE, RAE and constraint (i.e., all
  tuples with the components of a candidate rule other than the operation
  set), and the number of such tuples is very large.  In contrast, our
  algorithm never enumerates such candidates.} %
\fullonly{Molloy {\it et al.} apply a machine learning algorithm that uses
  a statistical approach, based upon a generative model, to find the RBAC
  policy that is most likely to generate the behavior (usage of
  permissions) observed in the logs \cite{molloy12generative}.
  They give an algorithm, based on Rosen-Zvi et al.'s algorithm for
  learning Author-Topic Models (ATMs) \cite{rosenzvi10learning}, to mine
  meaningful roles from logs and attribute data, i.e., roles such that the
  user-role assignment is statistically correlated with user attributes.  A
  discretization algorithm based on simulated annealing is used to produce
  an RBAC policy from the probabilistic results of the 
  ATM algorithm.  We adapted their approach to ABAC policy mining from logs
  as described in Section \ref{sec:experiments:atm}, and found that it is
  less effective than our algorithm on the small examples we tried.
  Another issue with this approach is scalability: the adapted algorithm
  enumerates all authors, and the number of authors is very large, because
  authors correspond to tuples containing a well-formed UAE, RAE and
  constraint (i.e., tuples with the components of a candidate rule other
  than the operation set).
  In contrast, our algorithm never enumerates all well-formed candidate
  rules; it constructs candidate rules by generalizing specific rules
  constructed from seeds in the input.  The number of authors is much
  smaller in Molloy {\it et al.}'s work, which does not consider resource
  attributes, and hence an author corresponds to only a candidate user
  attribute expression.  Another issue with this approach is that it
  requires specifying the desired number of rules, which is difficult to
  predict.}

\shortonly{ Zhang {\it et al.}  use machine learning algorithms to improve
  the quality of a given role hierarchy based on users' access patterns as
  reflected in operation logs \cite{zhang11role,zhang13evolving}.  These
  papers do not consider improvement or mining of ABAC policies. }%
\fullonly{ Zhang {\it et al.}  apply the experience-based access management
  approach \cite{gunter11experience} for role evolution \cite{zhang11role}.
  In particular, they use machine learning algorithms based on N\"aive
  Bayes classifiers to assess the quality of the roles in a given role
  hierarchy relative to given access logs.  The quality is based on the
  strength of the correlation between the user-role assignment and users'
  access patterns.  The analysis results can be used to improve the role
  hierarchy.  Zhang {\it et al.} also propose an algorithm based on Support
  Vector Machines to evolve a role hierarchy \cite{zhang13evolving}.  Given
  a role hierarchy and access logs, the algorithm tries to generate a new
  role hierarchy that optimizes a policy quality metric based on both the
  distance between the given policy and the generated policy and the
  homogeneity of the generated roles with respect to the logs.  Role
  homogeneity is a measure of the similarity of the access patterns of the
  members of a role.  These papers do not consider improvement or mining of
  ABAC policies.  }


\fullonly{\section{Conclusion}
\label{sec:conclusion}

This paper presents an algorithm for mining ABAC policies from logs and
attribute data.  Experiments with case studies and synthetic policies
demonstrate the algorithm's effectiveness even when the log reflects only a
fraction of the entitlements.  Although the original (desired) ABAC policy
is not reconstructed perfectly from the log, the mined policy is
sufficiently similar to it that the mined policy would be very useful as a
starting point for policy administrators tasked with developing that ABAC
policy.

Directions for future work include\fullonly{ handling additional policy
  language constructs such as arithmetic inequalities and negation,} better
automated tuning of parameters such as $w'_o$, characterization of the
algorithm's effectiveness as a function of input characteristics such as
the ratio of frequencies of the most and least frequent items ({\it cf.}
Section \ref{sec:log-gen}) and the complexity of the ABAC policy (average
number of conjuncts per rule, etc.), and experiments with noisy logs.
}








\fullonly{\myparagraph{Acknowledgments}
Christian Hesselbach did the comparison with Progol.
}

\shortonly{\vspace{-1.7ex}}
\lncsonly{\bibliographystyle{splncs03}}\articleonly{\bibliographystyle{alpha}}
\bibliography{../references}

\fullonly{
\appendix
\section{Rule Quality Metric Based On Inductive Logic Programming}
\label{sec:rule-quality-ILP}

We review the theory quality metric used in Progol
\cite{muggleton00theory,muggleton01cprogol}, a well-known ILP system, and
then describe our design of a rule quality metric based on it.  Progol's
IPL algorithm works as follows. At the outermost level, Progol uses a loop
that repeatedly generalizes an example to a hypothesized rule and then
removes examples which are redundant relative to (i.e., covered by) the new
rule, until no examples remain to be generalised. When generalizing an
example, Progol uses a metric, called a {\em compression metric}, to guide
construction of the hypotheses.  When mining ABAC policies from operation
logs, user-permissions tuples in $\up(L)$ are positive examples, and no
negative examples are available.  Thus, this corresponds to the case of
learning from only positive data.  When learning from only positive data,
Progol's compression metric $\pcomp$ is defined as follows
\cite{muggleton95inverse}.
\begin{eqnarray}
  f_m(H) &=& c\times 2^{-|H|}(1-g(H))^m \label{eq:fmdef}\\
  \pcomp(H,E) & = & \log_2\frac{f_m(H)}{f_m(E)} \label{eq:pcomp}\\
  & = & 
  \begin{array}[t]{@{}l@{}}
    |E| - |H| - m(\log_2(1-g(E))\\
    {} - \log_2(1-g(H)))  
  \end{array}\nonumber\\
  & \approx & |E| - |H| + m\log_2(1-g(H))\nonumber
\end{eqnarray}
where $E$ is the set of positive examples, $H$ is the entire theory (i.e.,
ABAC policy, in our context) being generated, including the part not
generated yet, $m = |E|$, $|H|$ is the size of $H$, measured as the number
of bits needed to encode $H$, and $c$ is a constant chosen so that
$\sum_{H\in {\cal H}}f_m(H) = 1$, where ${\cal H}$ is the set of all
candidate theories (note that $f$, like $\pcomp$ is a function of $H$ and
$E$, since $m=|E|$, but we adopt Muggleton's notation of $f_m(H)$ instead
of using the more straightforward notation $f(H,E)$).  Let $X$ be the set
of all possible well-formed examples (in our context, $X$ is the set of all
user-permission tuples).  $g(H)$ (the "generality" of $H$) is the
probability that an element of $X$, randomly selected following a uniform
distribution, satisfies $H$.  $f_m$ and $\pcomp$ can be regarded as policy
quality metrics.  The term $2^{-|H|}$ in the definition of $f_m$ causes
policies with smaller size to have higher quality, and the term
$(1-g(H))^m$ causes policies with larger meaning to have higher quality.
In \cite{muggleton95inverse}, $f_m$ is used to guide the search for rules.
Using $\pcomp$ to guide the search would have the same effect, because in
the definition of $\pcomp$ in equation (\ref{eq:pcomp}), $f_m(E)$ is a
constant, and $\log_2\frac{f_m(H)}{f_m(E)}$ is a monotonic function of
$f_m(H)$, so maximizing $f_m$ is equivalent to maximizing $\pcomp$.

A difficulty with using $f_m$ to guide generation of a rule to add to a
partly generated theory is that the entire theory $H$ is not yet known.
To overcome this difficulty, the quality of the entire theory is estimated
by extrapolation.  Let $C_i$ denote the $i$'th rule added to the theory,
and let $H_i = \set{C_1, \ldots, C_i}$.  Let $n$ denote the number of rules
in the entire theory being generated (of course, $n$ is not known until the
algorithm terminates).  When generating $C_i$, the policy quality
$f_m(H_n)$ is estimated as follows.
\begin{equation}
  \label{eq:fm-estimate}
  c\times 2^{-(\frac{m}{p}\times|C_i|)}\times(1 - \frac{m}{p}*(g(H_i) - g(H_{i-1})))^m
\end{equation}
where $p$ is the number of examples in $E$ that are implied by $C_i$ and
not by $H_{i-1}$.

Now we describe how to modify our policy mining algorithm to use $f_m$ as a
rule quality metric.  In the loop in the top-level pseudocode in Figure
\ref{fig:alg} that builds the set $\Rho$ of candidate rules, rule quality
is computed using equation \ref{eq:fm-estimate} with $H_{i-1}=\Rho$ and
$C_i=\rho$.  Specifically, in the definition of $\generalizeruleL$ in
\ref{fig:generalizerule}, $\Qrul(\rho,\uncovup$) is replaced with $f_m$
computed using equation \ref{eq:fm-estimate} with $H_{i-1} = \Rho$ and $C_i
= \rho$.  This closely corresponds to the usage in Progol, although it is
slightly different, because some of the candidate rules in $\Rho$ will not
be included in the final set of rules $\Rho'$.

In the loop in the top-level pseudocode in Figure \ref{fig:alg} that builds
the set $\Rho'$ of final rules, we take the same approach, except using
$\Rho'$ instead of $\Rho$.  Specifically, $\Qrul(\rho,\uncovup)$ is
replaced with $f_m$ computed using equation \ref{eq:fm-estimate} with
$H_{i-1}=\Rho'$ and $C_i=\rho$.

To compute rule quality after building the set $\Rho$ of candidate rules
and before building the set $\Rho'$ of final rules, the algorithm is
modifying a set of rules, not extending a set of rules, so $f_m$ can be
evaluated using equation \ref{eq:fmdef}, with $\Rho$ as an estimate of the
entire policy $H$.  Specifically, in the calls to $\elimconjunctsL$ and
$\elimconstraintsL$ from $\simplifyrulesL$, $\Qrul(\rho'',\up_0)$ is replaced
with $f_m$ computed using equation \ref{eq:fmdef} with $H = \Rho \setminus
{\rho} \union \{\rho''\}$, and $\Qrul(\rhobest,\up_0)$ is replaced with
$f_m$ computed using equation \ref{eq:fmdef} with $H = \Rho \setminus
{\rho} \union \{\rhobest\}$.



}

\end{document}
